\documentclass[10pt]{article}

\usepackage{graphicx}

\usepackage{amsfonts}

\usepackage{theorem}

\theorembodyfont{\upshape} \newtheorem{conjecture}{Conjecture}

\newcommand{\be}{\begin{equation}} \newcommand{\ee}{\end{equation}}
\newcommand{\bea}{\begin{eqnarray}} \newcommand{\eea}{\end{eqnarray}}
\newcommand{\nn}{\nonumber \\} 
 \newcommand{\tab}{Table~\ref}
\newcommand{\sect}{Section~\ref} \newcommand{\eqn}{Eq.~\ref}

\newlength{\diaght} 

\newlength{\diagshift} 

\newcommand{\mayeriva}{\raisebox{\diagshift}{\includegraphics[height=\diaght]{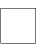}}}
\newcommand{\mayerivb}{\raisebox{\diagshift}{\includegraphics[height=\diaght]{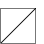}}}
\newcommand{\mayerivc}{\raisebox{\diagshift}{\includegraphics[height=\diaght]{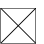}}}
\newcommand{\insiva}{\raisebox{\diagshift}{\includegraphics[height=\diaght]{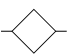}}}
\newcommand{\insivb}{\raisebox{\diagshift}{\includegraphics[height=\diaght]{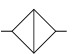}}}
\newcommand{\insva}{\raisebox{\diagshift}{\includegraphics[height=\diaght]{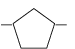}}}
\newcommand{\insvb}{\raisebox{\diagshift}{\includegraphics[height=\diaght]{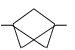}}}
\newcommand{\insvc}{\raisebox{\diagshift}{\includegraphics[height=\diaght]{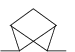}}}
\newcommand{\insvd}{\raisebox{\diagshift}{\includegraphics[height=\diaght]{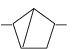}}}
\newcommand{\insve}{\raisebox{\diagshift}{\includegraphics[height=\diaght]{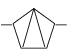}}}
\newcommand{\insvf}{\raisebox{\diagshift}{\includegraphics[height=\diaght]{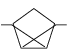}}}
\newcommand{\insvg}{\raisebox{\diagshift}{\includegraphics[height=\diaght]{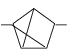}}}
\newcommand{\insvh}{\raisebox{\diagshift}{\includegraphics[height=\diaght]{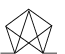}}}
\newcommand{\insvi}{\raisebox{\diagshift}{\includegraphics[height=\diaght]{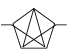}}}
\newcommand{\wigglya}{\ensuremath{\emptyset}}
\newcommand{\wigglyb}{\raisebox{\diagshift}{\includegraphics[height=\diaght]{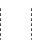}}}
\newcommand{\wigglyc}{\raisebox{\diagshift}{\includegraphics[height=\diaght]{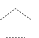}}}
\newcommand{\wigglyd}{\raisebox{\diagshift}{\includegraphics[height=\diaght]{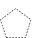}}}
\newcommand{\wigglye}{\raisebox{\diagshift}{\includegraphics[height=\diaght]{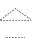}}}
\newcommand{\wigglyf}{\raisebox{\diagshift}{\includegraphics[height=\diaght]{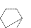}}}
\newcommand{\wigglyg}{\raisebox{\diagshift}{\includegraphics[height=\diaght]{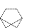}}}
\newcommand{\wigglyh}{\raisebox{\diagshift}{\includegraphics[height=\diaght]{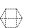}}}
\newcommand{\wigglyi}{\raisebox{\diagshift}{\includegraphics[height=\diaght]{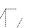}}}
\newcommand{\wigglyj}{\raisebox{\diagshift}{\includegraphics[height=\diaght]{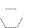}}}
\newcommand{\wigglyk}{\raisebox{\diagshift}{\includegraphics[height=\diaght]{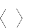}}}
\newcommand{\wigglyl}{\raisebox{\diagshift}{\includegraphics[height=\diaght]{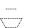}}}
\newcommand{\wigglym}{\raisebox{\diagshift}{\includegraphics[height=\diaght]{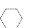}}}
\newcommand{\wigglyn}{\raisebox{\diagshift}{\includegraphics[height=\diaght]{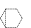}}}
\newcommand{\wigglyo}{\raisebox{\diagshift}{\includegraphics[height=\diaght]{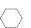}}}
\newcommand{\wigglyp}{\raisebox{\diagshift}{\includegraphics[height=\diaght]{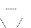}}}
\newcommand{\wigglyq}{\raisebox{\diagshift}{\includegraphics[height=\diaght]{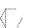}}}
\newcommand{\wigglyr}{\raisebox{\diagshift}{\includegraphics[height=\diaght]{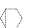}}}
\newcommand{\wigglys}{\raisebox{\diagshift}{\includegraphics[height=\diaght]{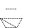}}}
\newcommand{\wigglyt}{\raisebox{\diagshift}{\includegraphics[height=\diaght]{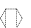}}}
\newcommand{\wigglyu}{\raisebox{\diagshift}{\includegraphics[height=\diaght]{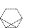}}}
\newcommand{\wigglyv}{\raisebox{\diagshift}{\includegraphics[height=\diaght]{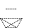}}}
\newcommand{\wigglyw}{\raisebox{\diagshift}{\includegraphics[height=\diaght]{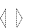}}}

\newcommand{\rhva}{\raisebox{\diagshift}{\includegraphics[height=\diaght]{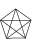}}}
\newcommand{\rhvb}{\raisebox{\diagshift}{\includegraphics[height=\diaght]{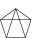}}}
\newcommand{\rhvc}{\raisebox{\diagshift}{\includegraphics[height=\diaght]{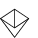}}}
\newcommand{\rhvd}{\raisebox{\diagshift}{\includegraphics[height=\diaght]{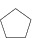}}}
\newcommand{\rhve}{\raisebox{\diagshift}{\includegraphics[height=\diaght]{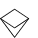}}}
\newcommand{\rhvia}{\raisebox{\diagshift}{\includegraphics[height=\diaght]{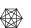}}}
\newcommand{\rhvib}{\raisebox{\diagshift}{\includegraphics[height=\diaght]{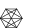}}}
\newcommand{\rhvic}{\raisebox{\diagshift}{\includegraphics[height=\diaght]{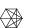}}}
\newcommand{\rhvid}{\raisebox{\diagshift}{\includegraphics[height=\diaght]{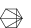}}}
\newcommand{\rhvie}{\raisebox{\diagshift}{\includegraphics[height=\diaght]{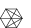}}}
\newcommand{\rhvif}{\raisebox{\diagshift}{\includegraphics[width=\diaght,angle=-90,origin=c]{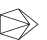}}}
\newcommand{\rhvig}{\raisebox{\diagshift}{\includegraphics[height=\diaght]{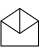}}}
\newcommand{\rhvih}{\raisebox{\diagshift}{\includegraphics[height=\diaght]{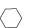}}}
\newcommand{\rhvii}{\raisebox{\diagshift}{\includegraphics[width=\diaght,angle=90,origin=c]{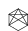}}}
\newcommand{\rhvij}{\raisebox{\diagshift}{\includegraphics[height=\diaght]{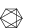}}}
\newcommand{\rhvik}{\raisebox{\diagshift}{\includegraphics[height=\diaght]{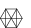}}}
\newcommand{\rhvil}{\raisebox{\diagshift}{\includegraphics[height=\diaght]{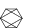}}}
\newcommand{\rhvim}{\raisebox{\diagshift}{\includegraphics[height=\diaght]{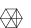}}}
\newcommand{\rhvin}{\raisebox{\diagshift}{\includegraphics[height=\diaght]{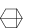}}}
\newcommand{\rhvio}{\raisebox{\diagshift}{\includegraphics[height=\diaght]{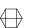}}}
\newcommand{\rhvip}{\raisebox{\diagshift}{\includegraphics[height=\diaght]{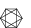}}}
\newcommand{\rhviq}{\raisebox{\diagshift}{\includegraphics[width=\diaght,angle=90,origin=c]{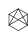}}}
\newcommand{\rhvir}{\raisebox{\diagshift}{\includegraphics[height=\diaght]{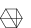}}}
\newcommand{\rhvis}{\raisebox{\diagshift}{\includegraphics[height=\diaght]{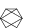}}}
\newcommand{\rhvit}{\raisebox{\diagshift}{\includegraphics[height=\diaght]{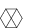}}}
\newcommand{\rhviu}{\raisebox{\diagshift}{\includegraphics[height=\diaght]{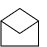}}}
\newcommand{\rhviv}{\raisebox{\diagshift}{\includegraphics[height=\diaght]{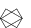}}}
\newcommand{\rhviw}{\raisebox{\diagshift}{\includegraphics[height=\diaght]{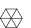}}}

\begin{document}

\title{Negative virial coefficients and the dominance of loose packed
diagrams for $D$--dimensional hard spheres}

\author{N.~Clisby\footnote{C.~N.~Yang Institute for Theoretical Physics, State
University of New York at Stony Brook, Stony Brook, NY 11794-3840; e-mail: \nobreak{Nathan.Clisby@stonybrook.edu} and \nobreak{mccoy@insti.physics.sunysb.edu}} \ and B.~M.~McCoy$^*$}

\maketitle

\begin{abstract}
We study the virial coefficients $B_k$ of hard spheres in $D$
dimensions by means of Monte-Carlo integration. We find that $B_5$ is
positive in all dimensions but that $B_6$ is negative for all $D\geq
6$. For $7\leq k\leq 17$ we compute sets of Ree-Hoover diagrams and
find that either for large $D$ or large $k$ the dominant diagrams are
``loose packed''.  We use these results to study the radius of
convergence and the validity of the many approximations used for the
equations of state for hard spheres.
\end{abstract}



\section{Introduction}
\label{introsec}

The question of the possible negativity of virial coefficients $B_k$
in the low density expansion \be \frac{P}{k_BT}=
\rho+\sum_{k=2}^{\infty}B_k \rho^{k}
\label{vir}
\ee of the system of hard spheres with diameter $\sigma$ in $D$
dimensions specified by the two body pair potential \be U({\bf r}) =
\left\{ \begin{array}{c} +\infty \\ 0 \end{array}
\right. \begin{array}{c} |{\bf r}|<\sigma \\ |{\bf r}|> \sigma \\
\end{array}
\label{hard}
\ee has been an unresolved problem of outstanding importance since it
was first proposed by Temperley~\cite{temperley1957a} in 1957. For
dimensions $D\leq 5$ all currently available information is summarized
in \tab{virialtable} where it is seen that all $B_k$ are
positive. However it was first shown~\cite{ree1964b} in 1964 for
$D\geq 8$ that $B_4$ is indeed negative. The best available current
results for $B_4$ as a function of dimension are shown in
\tab{analyticalB4table}.


\begin{table}[!hbt]
\caption{Numerical values of the virial coefficients $B_k/B_2^{k-1}$
for $k=3,\ldots, 8$ for $D=2,3,4,5$.}
\label{virialtable}
\vspace{0.2cm}
\begin{center}
\begin{tabular}{|l|l|l|}\hline
&discs&spheres\\ \hline $B_2$&$\pi \sigma^2/2$&$2\pi
\sigma^3/3$\\
$B_3/B_2^2$&$0.782004\cdots$\cite{tonks1936a}
&$0.625$\cite{boltzmann1899a}
\\
$B_4/B_2^3$&$0.5322318\cdots$\cite{rowlinson1964a,hemmer1964a}
&$0.2869495\cdots$\cite{boltzmann1899a,vanlaar1899a,nijboer1952a}
\\
$B_5/B_2^4$&$0.33355604(4)$\cite{ree1964a,kratky1982a}
&$0.110252(1)$\cite{ree1964a,kratky1977a}\\
$B_6/B_2^5$&$0.19883(1)$\cite{ree1964a,kratky1982c}
&$0.038808(55)$\cite{ree1964a,vanrensburg1993a}
\\
$B_7/B_2^6$&$0.114877(11)$\cite{ree1967a,vanrensburg1993a,vlasov2002a}
&$0.013046(22)$\cite{ree1967a,vanrensburg1993a,vlasov2002a}\\
$B_8/B_2^7$&$0.065030(31)$\cite{vanrensburg1993a,vlasov2002a}
&$0.004164(16)$\cite{vanrensburg1993a,vlasov2002a}\\ 
\hline
\hline
&$D=4$&$D=5$\\ \hline $B_2$&$\pi^2 \sigma^4/4$&$4\pi^2 \sigma^5/15$\\
$B_3/B_2^2$&$0.50634\cdots$\cite{luban1982a}&$0.414062\cdots$\cite{luban1982a}\\
$B_4/B_2^3$&$0.15184606\cdots$\cite{clisby2003a}&$0.075972512(4)$\cite{bishop1999a}\\
$B_5/B_2^4$&$0.03563(7)$\cite{bishop1999a}
&$0.01287(6)$\cite{bishop1999a}\\
$B_6/B_2^5$&$0.007691(28)$\cite{bishop1999a}&$0.000942(27)$\cite{bishop1999a}\\
\hline
\end{tabular}
\normalsize
\end{center}
\end{table}

\renewcommand{\arraystretch}{1.3}
\begin{table}[!hbt]
\caption{Exact and numerical results for $B_2, B_3$, and $B_4$ for
$2\leq D\leq 12$.}
\label{analyticalB4table}
\small
\vspace{0.1cm}
\begin{center}
\begin{tabular}{|l|l|l|l|l|}
\hline $D$ & {$B_2$} & {$B_3/B_2^2$} &\multicolumn{2}{c|}{$B_4/B_2^3$ exact and numerical} \\ \hline 2&$\frac{\pi
\sigma^2}{2}$&$\frac{4}{3}-{\frac{{\sqrt 3}}{
\pi}}$&$2-\frac{9\sqrt{3}}{2\pi}+\frac{10}{\pi^2}$&$\;\;\>0.53223180\cdots$\\
3&$\frac{2\pi\sigma^3}{3}$&$5/8$ &$\frac{219\sqrt
2}{2240\pi}-\frac{89}{280}+\frac{4131}{ 2240\pi}{\rm arctan}{\sqrt
2}$&$\;\;\>0.2869495\cdots$\\ 4
&$\frac{\pi^2\sigma^4}{4}$&$\frac{4}{3}-{\frac{\sqrt{3}}{\pi}}\frac{3}{2}$
&$2-\frac{27\sqrt{3}}{4\pi}+\frac{832}{45\pi^2}$\cite{clisby2003a}
&$\;\;\>0.15184606\cdots$\\ 5& $\frac{4\pi^2\sigma^5}{15}$&$ 53/2^7$&
&$\;\;\>0.07597$\cite{bishop1999a} \\
6&$\frac{\pi^3\sigma^6}{12}$&$\frac{4}{3}-{\frac{\sqrt{3}}{\pi}}{\frac{9}{5}}$
&$2-\frac{81\sqrt{3}}{10\pi}+\frac{38848}{1575\pi^2}$\cite{clisby2003a}
&$\;\;\>0.03336314\cdots$\\ 7& $\frac{8\pi^3\sigma^7}{105}$ & $289/2^{10}$&
&$\;\;\>0.0098$\cite{ree1964b} \\ 8
&$\frac{\pi^4\sigma^8}{945}$&$\frac{4}{3}-{\frac{\sqrt{3}}{\pi}}{\frac{279}{140}}$
&$2-\frac{2511\sqrt{3}}{280\pi}+\frac{17605024}{606375\pi^2}$\cite{clisby2003a}
&$-0.00255768\cdots$\\ 9&$\frac{16\pi^4\sigma^9}{48}$ &
$6343/2^{15}$&&$-0.00841$\cite{ree1964b} \\ 10
&$\frac{\pi^5\sigma^{10}}{240}$&$\frac{4}{3}-{\frac{\sqrt{3}}{\pi}}{\frac{297}{140}}$
&$2-\frac{2673\sqrt{3}}{280\pi}+\frac{49048616}{1528065\pi^2}
$\cite{clisby2003a}&$-0.01096248\cdots$\\
11&$\frac{32\pi^5\sigma^{11}}{10395}$&$35995/2^{18}$& &\\ 12&$\frac{\pi^6
\sigma^{12}}{1440}$&$\frac{4}{3}-{\frac{\sqrt{3}}{\pi}}{\frac{243}{110}}$
&$2-\frac{2187\sqrt{3}}{220\pi}+\frac{11565604768}{337702365\pi^2}$
\cite{clisby2003a}&$-0.01067028\cdots$\\ \hline
\end{tabular}
\end{center}
\normalsize
\end{table}
\renewcommand{\arraystretch}{1}

Most of our intuition and physical insight into the low density
(fluid) phase of hard spheres in 3 dimensions comes from the 8 term
virial expansion of \tab{virialtable}. Over the years this data has
been used to produce many approximate equations of
state~\cite{ree1964a,vanrensburg1993a,thiele1963a,wertheim1963a,
wertheim1964a,guggenheim1965a,carnahan1969a,
hoover1968a,hoste1984a,goldman1988a,
lefevre1972a,ma1986a,jasty1987a,song1988a,wang2002a,torquato1995a,torquato1995b}.  These
approximates all incorporate the feature of positive virial
coefficients and they all have the feature that they have a radius of
convergence which is greater than the packing fraction $\eta_f=0.49$
at which freezing has been seen to occur in computer
experiments~\cite{alder1957a,alder1960a,hoover1968a} where the packing
fraction $\eta$ is related to the density $\rho$ by $\eta=B_2
\rho/2^{D-1}.$ This analyticity at the freezing density is
incorporated into most phenomenological theories of
freezing~\cite{kirkwood1940a,ramakrishnan1979a,laird1987a} as a
homogeneity or mean field approximation which ignores the fluctuations
at phase coexistence between the fluid and solid
phases~\cite{haymet1986a}.

However, if there are negative virial coefficients for hard spheres in
$D=3$ then no conclusion on the radius of convergence of the virial
expansion based on \tab{virialtable} can be considered as reliable.

The most striking effect of negative viral coefficients will occur if
the signs oscillate with some period as $k\rightarrow\infty$ because
this will give a radius of convergence which is not on the positive
real axis. If this radius is less than the freezing density then it
will be impossible to reliably learn anything about the freezing
transition from a knowledge of a finite number of virial coefficients.

It is thus most significant that the sixth and seventh virial
coefficients for parallel hard cubes~\cite{hoover1962a} were shown to
be negative in 1962, and even more important that for an exactly
solved hard squares model~\cite{baxter1970a} and the hard hexagon
model~\cite{baxter1980a,richey1987a,joyce1988a} the radius of
convergence is limited by a singularity at complex density thus
resulting in virial coefficients that oscillate in sign.

The purpose of this paper is to extend the results of Tables
\ref{virialtable} and \ref{analyticalB4table} and examine as closely
as possible the question of whether or not the virial coefficients
$B_k$ of the hard sphere gas in $D$ dimensions have negative virial
coefficients. The method we shall use is Monte-Carlo evaluations of
the integrals in the Ree-Hoover expansion. In \sect{reehoovsec} we
review the formalism of the Ree-Hoover expansion to establish our
notation. In \sect{montesec} we compute the virial coefficients $B_5$
and $B_6$ for dimensions up to $D=50$. We find that $B_5$ is not
monotonic but is in fact always positive. More importantly we find
that $B_6$ is negative for all $D\geq 6.$ Our numerical results are
given in \tab{B456table}. For higher virial coefficients the number of
contributing diagrams rapidly increases. Consequently in this study we
restrict our attention to various classes of diagrams which are
studied in \sect{traditionalsec} where we are able to determine the
class of diagrams which are dominant for large $k.$ In \sect{radiisec}
we use our results to form estimates of the radius of convergence to
the virial series and we conclude in \sect{eqstatesec} with an
evaluation of the various approximate equations of state for hard
spheres.

\begin{table}[!htb]
\caption{Numerical results for $B_4/B_2^3$, $B_5/B_2^4$, and
$B_6/B_2^5.$ The underline indicates the position of the local minima
and maxima. Values for each coefficient for $D=3$, and for $B_6$ in
$D=4,5$ are taken from \tab{virialtable}.}
\label{B456table}
\vspace{0.1cm}
\begin{center}
\begin{tabular}{|l|l|l|l|}
\hline $D$ & {$B_4/B_2^3$} & {$B_5/B_2^4$} & {$B_6/B_2^5$} \\ \hline 3
&$\;\;\> 0.2869495\cdots $ &$0.110252(1)$ &$\;\;\>0.03881(6)$ \\ 4 &$
\;\;\> 0.1518460\cdots $ & $ 0.03565(5) $ &$\;\;\>0.00769(3)$ \\ 5 &$
\;\;\> 0.075978(4) $ & $ 0.01297(1) $ & $ \; \; \> 0.00094(3) $ \\ 6
&$ \; \; \> 0.03336314\cdots $ & $ 0.007528(8) $ & $ -0.00176(2) $ \\
7 &$ \; \; \> 0.009873(4) $ & $ \underline{0.007071(7)} $ & $
-0.00352(2) $ \\ 8 &$-0.0025576\cdots$ & $ 0.007429(6) $ & $
-0.00451(2) $ \\ 9 &$-0.008575(3) $ & $\underline{ 0.007438(6)} $ & $
\underline{-0.00478(1)} $ \\ 10 &$-0.0109624\cdots$ & $ 0.006969(5) $
& $ -0.00452(1) $ \\ 11 &$\underline{-0.011334(3)} $ & $ 0.006176(4) $
& $ -0.00395(1) $ \\ 12 &$-0.0109624\cdots$ & $ 0.005244(4) $ & $
-0.003261(7) $ \\ 13 &$-0.009523(2) $ & $ 0.004307(3) $ & $
-0.002580(6) $ \\ 14 &$ -0.008220(2) $ & $ 0.003448(3) $ & $
-0.001975(4) $ \\ 15 &$ -0.006934(2) $ & $ 0.002705(2) $ & $
-0.001472(3) $ \\ 20 &$ -0.0024621(7) $ & $ 0.0006605(7) $ & $
-0.0002632(7) $ \\ 25 &$ -0.0007580(3) $ & $ 0.0001348(2) $ & $
-3.72(1) \times 10^{-5} $ \\ 30 &$ -0.0002196(1)$ & $ 2.515(6) \times
10^{-5} $ & $ -4.69(3) \times 10^{-6} $ \\ 35 &$-6.162(3) \times
10^{-5}$ & $ 4.47(1) \times 10^{-6} $ & $ -5.55(5) \times 10^{-7} $ \\
40 &$ -1.697(1) \times 10^{-5}$ & $ 7.69(3) \times 10^{-7} $ & $
-6.30(9) \times 10^{-8} $ \\ 45 &$ -4.618(4) \times 10^{-6}$ & $
1.298(7) \times 10^{-7} $ & $ -7.0(2) \times 10^{-9} $ \\ 50 &$
-1.247(1) \times 10^{-6}$ & $ 2.16(1) \times 10^{-8} $ & $ -7.6(2)
\times 10^{-10} $ \\ \hline
\end{tabular}
\end{center}
\normalsize
\end{table}

\section{Ree--Hoover expansions}
\label{reehoovsec}

The original graphical expansion for the virial series is due to Mayer
and Mayer~\cite{mayer1940a}, in which each bond represents the
function \be f({\bf r})=\exp\left(-U({\bf r})/{k_BT}\right)-1 \ee
where {\bf r} is the distance between the two vertices. A useful
re-summation was performed by Ree and Hoover~\cite{ree1964a,ree1964c}
by introducing the function \be {\tilde f(\bf r)}=1+f({\bf
r})=\exp\left(-U({\bf r})/{k_BT}\right) \ee and then expanding each
Mayer graph by substituting $1=\tilde{f}-f$ for pairs of vertices not
connected by $f$ bonds. This method was previously used by Percus and Yevick
\cite{percus1958a} in comparing the exact values of the fourth and fifth 
virial coefficients with coefficients obtained from the
Percus--Yevick equation, and by Percus \cite{percus1964a} in discussing
the derivation of the Percus--Yevick equation.

The fourth virial
coefficient may then be written as \be B_4 =-\frac{1}{8}
\,\mayerivc-\frac{3}{4} \, \mayerivb-\frac{3}{8} \, \mayeriva
=\frac{1}{4}\, \wigglya - \frac{3}{8} \, \wigglyb =\frac{1}{4}\,
\mayerivc - \frac{3}{8} \, \mayeriva
\label{b4rheq}
\ee where the first expression is the expansion in Mayer graphs, the
second is the expansion in Ree-Hoover graphs with the $\tilde f$ bonds
indicated by dotted lines and the third shows the equivalent
Ree-Hoover graphs with the $f$ bonds indicated by solid lines. In the
second expression the graph with no $\tilde f$ bonds is represented by
\wigglya.  In the case of hard spheres, the potential is given by
\eqn{hard} so $f({\bf r})$ and $\tilde{f}({\bf r})$ are particularly
simple:

\bea f({\bf r}) &=& \left\{ \begin{array}{r} -1 \\ 0
\end{array} \right. \begin{array}{l} |{\bf r}|< \sigma \\ |{\bf
r}|>\sigma\\ \end{array}
\label{fdef} \\ \tilde{f}({\bf r}) &=& \left\{ \begin{array}{r} 0 \\
+1 \end{array} \right. \begin{array}{l} |{\bf r}|<\sigma \\ |{\bf
r}|>\sigma \\
\end{array} \label{ftildedef} \eea

The virial coefficient $B_k$ is given in terms of the Ree-Hoover
diagrams of $k$ points of which $m$ are the end points of ${\tilde f}$
bonds $S_k[m,i]$ and a combinatorial factor $C_k[m,i]$ as \be
B_k=\sum_{m,i}B_k[m,i] \ee with \be B_k[m,i]=C_k[m,i]S_k[m,i] \ee
where the index $i$ labels the graphs in the class with fixed $k,m.$
The combinatorial factor is expressed as \be
C_k[m,i]=-\frac{k-1}{k!}s_{k}[m,i]{\tilde a}_k[m,i] \ee with \be
s_k[m,i]=k!/\sharp {\rm Aut} S_k[m,i] \ee where $\sharp {\rm Aut}
S_k[m,i]$ is the cardinality of the automorphism group of the diagram
$S_k[m,i]$ and $\tilde{a}_{m,l}[k]$ is the ``star content'' as defined
by Ree and Hoover~\cite{ree1964c}. We will let ${S}_k[m,i]$ represent
both the Ree-Hoover graph and the value of the corresponding
integral. The $k$ dependence of $C_k[m,i]$ is calculated by using the following
relation for the star content~\cite{ree1964c}.  \be \tilde{a}_k[m,i] =
(-1)^{k-1}(k-2)\tilde{a}_{k-1}[m,i]
\label{afactor}
\ee The diagram $S_k[m,i]$ has $k-m$ points that are connected to all
other points by $f$ bonds and are therefore indistinguishable, leading
to the relation \bea \sharp{\rm Aut} S_k[m,i] &=& (k-m)! \sharp{\rm
Aut} S_m[m,i]
\label{morefactor}
\eea to obtain for $k>m$ \be C_k[m,i] =(-1)^{k(k-1)/2} {k-1 \choose
m-1} C_m[m,i]
\label{comk}
\ee where we note that for the complete star diagram $B_k[0,1]$ we
have $C_k[0,1] =1/k.$

For arbitrary $D$ the number of Mayer graphs grows asymptotically as
$k\rightarrow \infty$ as~\cite{harary1970a} \be
\label{asymptoticgrowth}N(k) \sim\frac{2^{k(k-1)/2}}{k!} \ee However,
when the Ree-Hoover transformation is made many diagrams have zero
star content and hence do not contribute to the virial
coefficient. From \tab{reemayer} it appears that the ratio of
contributing Ree-Hoover diagrams to the number of Mayer diagrams is
bounded below, and so it is reasonable to suppose that \be \lim_{k\rightarrow \infty}N_{RH}(k)/N(k) > 0
\ee
\begin{table}[htb]
\caption{Number of contributing Ree-Hoover and Mayer diagrams as a
function of order. The number of Ree-Hoover diagrams with non-zero
star content for $k=9$ is a new result, while the other values are taken from \cite{ree1964a,ree1967a,vanrensburg1993a}.}
\label{reemayer}
\begin{center}
\begin{tabular}{|l|cccccccc|}
\hline &\multicolumn{8}{c|}{Order} \\ & $2$ & $3$ & $4$ & $5$ & $6$ &
$7$ & $8$ & $9$ \\ \hline Mayer&1 &1 &3 &10 &56 &468 &7123 & 194066\\
RH&1 &1 &2 &5 &23 &171 &2606 & 81564\\ RH, $D=2$ & 1 &1 &2 &4 &15
&78(66)& & \\ RH/Mayer& 1& 1& 0.667& 0.500& 0.410& 0.365& 0.366 &
0.420\\ \hline
\end{tabular}
\end{center}
\end{table}

In addition Ree-Hoover graphs may be zero for geometrical reasons. The
number of non-zero graphs for $D=2$ are taken from
\cite{ree1964a,ree1967a}, and listed in \tab{reemayer}, where the
value in parentheses excludes diagrams found to be negligible but
which were not proven to be zero.  For $D=1$ only one graph in the
Ree-Hoover expansion is non-zero for each $k$, namely the complete
star, but for $D\geq 2$ it is an open question as to how many non-zero
graphs there are at order $k$.

\subsection{Close packed diagrams}

The $\tilde{f}$ graph form used by Ree and Hoover (e.g. in
\cite{ree1964c}) has the property that a single graph represents the
sequence of diagrams $B_k[m,i]$ with $m$ fixed where the ($k+1$)th
order diagram is obtained from the $k$th by adding an additional point
connected to all other points by an $f$ bond.  For example, all
complete star diagrams of order $k$ may be represented by the same
symbol \wigglya.  We call such graphs with $m< k$ close packed because
no two points can be further apart than $2\sigma.$

We find from \eqn{comk} that the sign of $B_k[m,i]$ for fixed $m,i$ is
independent of $k.$ We further obtain from \eqn{comk} for $k$ large
and $m,i$ fixed that \be C_k[m,i] = C_{m}[m,i]\left(
\frac{k^{m-1}}{(m-1)!}+O(k^{m-2})\right) \ee

\subsection{Loose packed diagrams}

For diagrams in the class $B_k[k,i]$ there are no points which are
connected to all other points by $f$ bonds and therefore there exist
sequences of diagrams which have the property that the size of the
configuration grows as $k\rightarrow \infty.$ We refer to this class
of diagrams as ``loose packed''.  For these graphs the $f$ bond
notation is more convenient.

The simplest loose packed diagram in $B_k[k,i]$ is the simple ring of
$k$ of the $f$ bonds. We denote this diagram by the symbol ${\mathsf
R}$.  More generally we consider graphs where a point is replaced by a
diagram. We call such diagrams insertion diagrams and when we wish to
make the type of insertion visually apparent we use the notation
${\mathsf R}(\cdot)$ to represent a ring with a point replaced by a
diagram. We also find it useful to label insertion diagrams by
${\mathsf R}_{n,l}[k]$ where the index $n$ is the number of points in
the inserted diagram of $k$ total points and $l$ labels the diagrams
with given $k$ and $n.$ All four and five point insertions were found
by starting with all four and five point Mayer graphs, and adding an
extra point that is connected to two points by $f$ bonds. A canonical
labeling for this graph was found using ``nauty'', a program due to
B.~McKay~\cite{mckay1981a}, and matched with the five and six point
graphs of Ree and Hoover~\cite{ree1964a} to find the star content. All
insertions with non-zero star content are given in \tab{insertiontab}
along with the size of their automorphism group and over all
combinatorial factor.  From \eqn{comk} we find the insertion diagrams
have the alternating sign property that $(-1)^k {\mathsf R}_{n,l}[k]$
has a sign which depends only on $n$ and $l$ but is independent of
$k.$

Multiple insertions are also possible when $k$ is sufficiently
large. We denote such a graph with $n$ insertions as ${\mathsf
R}^{(m)}_{\{n_1,i_i\},\cdots,\{n_m,i_m\}}$ where the subscripts
indicate the types of insertions. For $B_6$ we see in \tab{b6rhtab}
that there are three graphs which may be interpreted as being composed
of two 4-point insertions.

\begin{table}[!hbt]
\caption{Four and Five point insertions}
\label{insertiontab}
\begin{center}
\begin{tabular}{|l|l|c|c|r|r|}
\hline Diagram & Label & Group size & Lowest order & $\tilde{a}_l[k]$
& $C_k[k,i]$ \\ \hline ${\mathsf R}$ & ${\mathsf R}$ &$2 k$ & 3 & 1 &
$-(k-1)/2k$\\ \hline ${\mathsf R}(\insiva)$ & ${\mathsf R}_{4,1}$& 4 &
6 & 1 & $-(k-1)/4$\\ ${\mathsf R}(\insivb)$ & ${\mathsf R}_{4,2}$& 4 &
5 & $-2$ & $(k-1)/2$\\ \hline ${\mathsf R}(\insva)$ & ${\mathsf
R}_{5,1}$& 2 & 8 & 1 & $-(k-1)/2$\\ ${\mathsf R}(\insvb)$ & ${\mathsf
R}_{5,2}$& 12 & 7 & 1 & $-(k-1)/12$\\ ${\mathsf R}(\insvc)$ &
${\mathsf R}_{5,3}$& 4 & 7 & $-1$ & $(k-1)/4$\\ ${\mathsf R}(\insvd)$
& ${\mathsf R}_{5,4}$& 1 & 7 & $-2$ & $2(k-1)$\\ ${\mathsf R}(\insve)$
& ${\mathsf R}_{5,5}$& 2 & 6 & 3 & $-3(k-1)/2$\\ ${\mathsf R}(\insvf)$
& ${\mathsf R}_{5,6}$& 4 & 7 & $-2$ & $(k-1)/2$\\ ${\mathsf
R}(\insvg)$ & ${\mathsf R}_{5,7}$& 2 & 6 & 1 & $(k-1)/2$\\ ${\mathsf
R}(\insvh)$ & ${\mathsf R}_{5,8}$& 4 & 6 & 3 & $-3(k-1)/4$\\ ${\mathsf
R}(\insvi)$ & ${\mathsf R}_{5,9}$& 12 & 6 & $-6$ & $(k-1)/2$\\ \hline
\end{tabular}
\end{center}
\end{table}

We give in Tables \ref{b4rhtab}--\ref{b6rhtab} the Ree-Hoover graphs
and their associated combinatorial factors for $k=4,5,6.$ We here give
both the representation of the graphs in terms of $\tilde f$ and $f$
bonds.  For the class of loose packed graphs in $B_k[k,i]$ we indicate
the interpretation in terms of insertion diagrams with either multiple
insertions or with a ring of one point and two bonds. In three cases
the identification is not unique.

The labeling index $i$ in $B_k[m,i]$ is chosen such that in $D=2$ the
magnitude of the contribution decreases with increasing $i$. If the
diagram is identically zero for $D=2$ then the ordering obtained from
$D=3$ is used when possible.

\renewcommand{\arraystretch}{1.5}
\begin{table}[!ht]
\caption{Ree-Hoover diagrams for $B_4$.}
\label{b4rhtab}
\begin{center}
\begin{tabular}{|c|r|r|r|c|c|c|}
\hline Label & $s_k[m,i]$ & $\tilde{a}_k[m,i]$ &$C_k[m,i]$&
$\tilde{f}$ form & $f$ form & Insertion \\ \hline
$B_4[0,1]$& 1 & $-2$ &$2/8$ &\wigglya & \mayerivc & \\ \hline
$B_4[4,1]$& 3 & 1 &$-3/8$ &\wigglyb & \mayeriva & ${\mathsf R}[4]$\\
\hline
\end{tabular}
\end{center}
\end{table}
\renewcommand{\arraystretch}{1}

\renewcommand{\arraystretch}{1.5}
\begin{table}[!ht]
\caption{Ree-Hoover diagrams for $B_5$.}
\label{b5rhtab}
\begin{center}
\begin{tabular}{|c|r|r|r|c|c|c|}
\hline Label & $s_k[m,i]$ & $\tilde{a}_k[m,i]$&$C_k[m,i]$ &$\tilde{f}$
form & $f$ form & Insertion\\ \hline $B_5[0,1]$& 1 &
$-6$ &6/30&\wigglya & \rhva & \\ \hline $B_5[4,1]$& 15 & 3 &$-45/30$&
\wigglyb & \rhvb & \\ \hline $B_5[5,1]$& 30 & $-2$ &60/30&\wigglyc &
\rhvc & ${\mathsf R}_{4,2}[5]$\\ $B_5[5,2]$& 12 & 1 &$-12/30$&
\wigglyd & \rhvd &${\mathsf R}[5]$\\ $B_5[5,3]$& 10 & 1
&$-10/30$&\wigglye & \rhve &${\mathsf R}_{4,1}[5]$\\ \hline
\end{tabular}
\end{center}
\end{table}
\renewcommand{\arraystretch}{1}

\renewcommand{\arraystretch}{1.5}
\begin{table*}[!tbh]
\caption{Ree-Hoover diagrams for $B_6$. For diagrams $B_6[6,14]$,
$B_6[6,16]$, and $B_6[6,17]$ the assignment of insertion diagram
labels is not unique and both possible assignments are shown.}
\label{b6rhtab}
\small
\begin{center}
\begin{tabular}{|c|r|r|r|c|c|c|}
\hline Label & {\footnotesize $s_k[m,i]$} & {\footnotesize $\tilde{a}_k[m,i]$}&$C_k[m,i]$ &
$\tilde{f}$ form & $f$ form & Insertion\\ \hline
$B_6[0,1]$& 1 & 24 &$-24/144$& \wigglya & \rhvia & \\ \hline
$B_6[4,1]$& 45 & $-12$ &$540/144$& \wigglyb & \rhvib & \\ \hline
$B_6[5,1]$& 180 & 8 &$-1440/144$& \wigglyc & \rhvic & \\ $B_6[5,2]$&
72 & $-4$ &$288/144$& \wigglyd & \rhvid & \\ $B_6[5,3]$& 60 &
$-4$&$240/144$ & \wigglye & \rhvie & \\ \hline $B_6[6,1]$& 360 & 3
&$-1080/144$& \wigglyf & \rhvif &${\mathsf R}_{5,5}[6]$\\ $B_6[6,2]$&
180 & $-2$& $360/144$& \wigglyg & \rhvig&${\mathsf R}_{4,2}[6]$ \\
$B_6[6,3]$& 60 & 1 &$-60/144$& \wigglyh & \rhvih&${\mathsf R}[6]$ \\
$B_6[6,4]$& 60 & $-6$&$360/144$ &\wigglyi & \rhvii&${\mathsf
R}_{5,9}[6]$ \\ $B_6[6,5]$& 180 & $-5$ &$900/144$& \wigglyj & \rhvij&
\\ $B_6[6,6]$& 90 & $-4$ &$360/144$& \wigglyk & \rhvik& \\ $B_6[6,7]$&
45 & 4 &$-180/144$& \wigglyl & \rhvil&${\mathsf
R}^{(2)}_{\{4,2\}\{4,2\}}[6]$ \\ $B_6[6,8]$& 360 & $-1$ &$360/144$&
\wigglym & \rhvim& \\ $B_6[6,9]$& 360 & $-2$ &$720/144$& \wigglyn &
\rhvin&${\mathsf R}_{5,4}[6]$ \\ $B_6[6,10]$& 60 & 4 &$-240/144$&
\wigglyo & \rhvio& \\ $B_6[6,11]$& 15 & 16 &$-240/144$& \wigglyp &
\rhvip& \\ $B_6[6,12]$& 180 & 3 &$-540/144$& \wigglyq &
\rhviq&${\mathsf R}_{5,8}[6]$ \\ $B_6[6,13]$& 360 & 1 &$360/144$&
\wigglyr & \rhvir&${\mathsf R}_{5,7}[6]$ \\ 
$B_6[6,14]$& 90 & $-2$
&$180/144$& \wigglys & \rhvis& ${\mathsf R}^{(2)}_{\{4,1\}\{4,2\}}[6]$ \\
&&&&&& or ${\mathsf R}_{5,6}[6]$ \\
 $B_6[6,15]$& 90 & $-1$ &$90/144$&
\wigglyt & \rhvit&${\mathsf R}_{5,3}[6]$ \\ $B_6[6,16]$& 180 & 1
&$-180/144$& \wigglyu & \rhviu& ${\mathsf R}_{4,1}[6]$ \\ &&&&&& or ${\mathsf
R}_{5,1}[6]$ \\ $B_6[6,17]$& 15 & 1 &$-15/144$& \wigglyv &
\rhviv&${\mathsf R}^{(2)}_{\{4,1\}\{4,1\}}[6]$ \\ &&&&&& or ${\mathsf
R}_{5,2}[6]$ \\ $B_6[6,18]$& 10 & 4 &$-40/144$& \wigglyw & \rhviw& \\
\hline
\end{tabular}
\end{center}
\normalsize
\end{table*}
\renewcommand{\arraystretch}{1}

\clearpage

\section{Monte-Carlo Calculation of $B_k$ for $D\ge k-1$.}
\label{montesec}

When $D \ge k-1$ the integral $S_k[m,i]$ can be written in a form
where the dimension $D$ appears as a simple power in the
integrand. Hence the Monte-Carlo procedure can simultaneously
calculate the given Ree-Hoover diagram in an arbitrary set of
dimensions, including non-integer values. The key advantage of this
method over that of \cite{ree1964a} used in \sect{traditionalsec} is
that we obtain fast convergence for high dimensions. We first present
the details of the Monte-Carlo method and then the results for $B_4$,
$B_5$, and $B_6$.

We believe that all Ree-Hoover diagrams are non-zero for $D \ge k-1$
as it is possible to obtain a configuration for any diagram. We can
see this by starting with a configuration where the distance between
each point is exactly one, and then since each bond can be
independently varied in length by a small amount we are able to
satisfy the constraints imposed by $f$ and $\tilde{f}$ bonds of any
graph.

\subsection{Monte-Carlo method}

In order to do the Monte-Carlo integration we need an appropriate
measure for calculating Ree-Hoover diagrams in $D$-dimensional
Euclidean space, where we wish to integrate out coordinates to leave
ourselves with the lowest dimensional integral possible. The integrand
is a product of $f$ and $\tilde{f}$ functions; for central potentials
this leaves only the inter-particle distances as appropriate degrees
of freedom. These form an independent set of $k(k-1)/2$ coordinates
provided that $D \ge k-1$, where $k$ is the order of the diagram and
hence the number of points in the configuration. After taking the
infinite volume limit, for an arbitrary diagram $S_k[m,i]$ we thus
need to calculate integrals of the form \be I = \int \prod_{i=1}^{k-1}
d^D{\bf r_i}{\bf F}( r_{ij})\nn = \left(\prod_{i=1}^{k-1}
\Omega_{D-i}\right) \int \prod_{i < j} da_{ij} [V(\{a_{ij}\})]^{D-k}
{\bf F}( r_{ij}) \label{distform} \ee where $\Omega_{D-1}\equiv{2\pi^{D/2} /
\Gamma(D/2)}$, $a_{ij} = |{\bf r_i}-{\bf r_j}|^2 = a_{ji}$, $V$ is the
volume of the parallel-piped defined by these distances in
$\mathbb{R}^{k-1}$, and ${\bf F}( r_{ij})$ is an arbitrary function of
the inter-particle distances. As shown for example
in \cite{blumenthal1953} $V$ may be expressed by
the Cayley-Menger determinant: \be
\label{simplex}
V(\{a_i\}) = \left[ \frac{(-1)^k}{2^{k-1}} \left|
\begin{array}{cccccc} 0 & 1 & 1 & 1 & \cdots & 1 \\ 1 & 0 & a_{12} &
a_{13} & \cdots & a_{1k} \\ 1 & a_{21} & 0 & a_{23} & \cdots & a_{2k}
\\ 1 & a_{31} & a_{32} & 0 & \cdots & a_{3k} \\ \vdots & \vdots &
\vdots & \vdots & \ddots & \vdots \\ 1 & a_{k1} & a_{k2} & a_{k3} &
\cdots & 0 \\ \end{array} \right| \right]^\frac{1}{2} \ee
An overall scale factor can be taken out of \eqn{simplex} by
taking $a_{12}$ to be the largest value, and then setting
$a_{ij}^\prime = a_{ij}/a_{12}$, so that \bea I &=& \frac{1}{2}k(k-1)
\left(\prod_{i=1}^{k-1} \Omega_{D-i}\right) \int_0^\infty da_{12}
a_{12}^{(k-1)D/2-1} \nn 
&& \times \left[ \prod_{i < j \neq 2}
\int_0^1 da_{ij}^\prime \right] V(\{a_{12}\equiv1,a_{ij}^\prime\})^{D-k} {\bf F}(r_{ij})
\label{measure}
\eea
A formula similar to \eqn{distform} was previously
obtained by Percus \cite{percus1987a}, in which the integration of $p$
vectors in dimension $D$ is reduced to an integration of $p$ vectors
in $p$ dimensions.

For the hard sphere fluid, the integrand is either zero or
$\pm 1$. In order to calculate $S_k[m,i]$
with $m$ $\tilde f$ bonds and $[k(k-1)/2-m]$ $f$ bonds, we may then
proceed as follows. Generate a set of $[k(k-1)/2-1]$ values uniformly
distributed between 0 and 1. Partition these values in two sets,
where the largest $m-1$ values along with 1 represent $\tilde f$ bonds
and the other
$[k(k-1)/2-m-1]$ values are $f$ bonds, and then randomly assign
these values to edges in the diagram. Check that these values of the
edge lengths squared can be embedded as a simplex in $\mathbb{R}^{k-1}$,
and if this is the case one can calculate the volume of the simplex
and perform the $a_{12}$ integral, making a contribution to the
Monte-Carlo integral of 
\be \frac{k}{D}
\left(a_{m}^{-(k-1)D/2}-a_{m+1}^{-(k-1)D/2}\right)
\left(\prod_{i=1}^{k-1} \Omega_{D-i}\right)
V(\{a_{12}\equiv1,a_{ij}^\prime\})^{D-k} \ee
where $a_m$ and $a_{m+1}$ are the $m$th and $(m+1)$th largest values of $a_{ij}$ respectively.

We have used  this procedure to compute $B_4$, $B_5$, and $B_6.$ 
For $B_4$ and $B_5$ 500
batches of $5\times10^6$ configurations were used, while for $B_6$ 20
batches of $2\times 10^9$ configurations were generated. Uncertainties were
calculated using
\be
\mathrm{Error} = \left[\sum_{j=1}^q
\frac{(<I_j>-<I>)^2}{q(q-1)}\right]^{\frac{1}{2}}
\label{montecarloerr}
\ee
where there are $q$ independent batches with value $I_j$.

\subsection{Results}

The values of the individual contributions $B_k[m,i]$ to the virial
coefficients $B_k$ for $k=4,5,6$ are given in Appendix \ref{inddiags} in  
Tables \ref{b4diagtab}--\ref{b6diagtab},
for many
values of $D\leq 50.$ The values of the virial coefficients $B_k/B_2^{k-1}$  
for $D\leq 50$ are given in \tab{B456table}.

The most important feature of the virial coefficients in 
\tab{B456table}
 are the sign changes in $B_4$ and $B_6.$ A previous estimate for the dimension 
at which $B_4$ becomes negative of $D\approx 7.8$ 
was obtained by Luban
and Baram~\cite{luban1982a} by means of a linear interpolation of the results 
of Ree and Hoover~\cite{ree1964b}, and also by an independent calculation.
We further note the existence of a local minimum in $B_5$  which 
confirms the tentative prediction of
Loeser et al.~\cite{loeser1993a}. To obtain accurate values for these
zeros, minima and maxima as a function of $D$ 
the data were fitted with cubic splines, and
all of these values are
summarized in \tab{maxmin}.
 Note that the error estimates were made by
fitting, for example, $B_4+\Delta B_4$ and $B_4-\Delta B_4$ with cubic
splines to obtain a confidence interval for where $B_4$ becomes
negative. A least squares fit was not used because the error in values
at different dimensions are not independent, as they are calculated
simultaneously and not as independent samples.

It is most important that $B_6$ becomes negative at a lower dimension
than $B_4.$ If this trend continues, we may expect that for
$B_{2k}$ the dimension at which the coefficient becomes negative will
decrease for higher orders and it is not unreasonable to expect that
$B_8$ will be negative for $D=5$ and possibly even $D=4.$
In addition, even though $B_5$ is always
positive, the existence of a local minimum at $D=6.87$ shows that the
increasing number of contributing diagrams results in more complex
dimensional dependence. 

\begin{table}[!tbh]
\caption{Maxima and minima for $B_4/B_2^3$, $B_5/B_2^4$, and
$B_6/B_2^5$.}
\label{maxmin}
\vspace{0.2cm}
\begin{center}
\begin{tabular}{|c|c|c|}
\hline $B_4/B_2^3$ & $B_5/B_2^4$ & $B_6/B_2^5$ \\ \hline Becomes
negative & Local minimum & Becomes negative\\ $D^{neg}_{4}=7.7320(4)$
& $D=6.87$ & $D^{neg}_6=5.30(2)$\\ Local minimum & Local maximum &
Local minimum\\ $D=10.7583(2)$ & $D=8.31$ & $D=8.942(2)$\\ \hline
\end{tabular}
\end{center}
\end{table}

To obtain further insight into the structure of the virial coefficients
we examine the individual contributions in Tables
\ref{b4diagtab}--\ref{b6diagtab}.
There we see by
examining the dependence of the contributions on $D$ that by the time $D=50$
the Ree-Hoover ring diagram ${\mathsf R}$ is several orders of magnitude
greater than all other diagrams. We therefore make the following conjecture:
\begin{conjecture}
\be
\lim _{D\rightarrow \infty} B_k(D)/{\mathsf R}(D)=1
\ee
\end{conjecture}
In particular it follows from this conjecture that for each $k$
\be
(-1)^{k-1}B_k(D)>0~~{\rm for~sufficiently~large}~ D.
\ee
A similar conjecture in terms of Mayer diagrams was made by Frisch and Percus \cite{frisch1999a} in the course of their examination of Mayer diagrams at high dimensionality.

We find further for $D=50$ that not only does the Ree-Hoover 
ring diagram dominate but that the largest diagram in each $B_k[m,i]$
has the property that 
if  $m>m'$ then
\be
B_k[m,i]_{\rm max}>>B_k[m',i]_{\rm max}.
\label{lpdom}
\ee
We refer to this ordering as the principle of ``loose packed dominance''.

We also note that within the loose packed class $B_k[k,i]$ that the
ordering of the magnitude of the diagrams at $D=2$ and at $D=50$ is
drastically different. For example we note that the diagram ${\mathsf R}_{4,1}$
which vanishes identically in $D=2$ is the second largest diagram for
$D=50.$ More generally we find for $D=50$ that the contributions in
$B_6[6,i]$ 
are ordered in magnitude as
\be
{\mathsf R}>{\mathsf R}_{4,1}>{\mathsf R}_{4,2}>{\mathsf R}_{5,4}>{\mathsf R}_{5,5}>
{\mathsf R}_{5,3}>{\mathsf R}_{5,2}={\mathsf R}^{(2)}_{\{4,1\},\{4,1\}}
\ee
where we use the labeling of insertion diagrams in 
Tables \ref{insertiontab} and \ref{b6rhtab}.
We make the observation that these insertion diagrams are the largest in
the class $B_k[k,i]$, and refer to this as ``insertion graph dominance''.

\section{Monte-Carlo Calculation of High Order Diagrams}
\label{traditionalsec}

The observation and conjecture of the dominance of the Ree-Hoover ring diagram
at large dimensions demonstrates that negative virial coefficients in
the hard sphere fluids is a common occurrence. However it says nothing
about the minimum dimension at which the coefficient $B_k$ will become
negative for fixed $k$ nor does it give any indication of whether or
not virial coefficients of odd order can ever be negative. Ideally
this question should be studied by computing further viral coefficients
beyond those in \tab{virialtable} for large $k.$ Here we will begin this 
program by examining selected classes of diagrams $B_k[m,i]$ for
values of $k$ up to 17. In particular we examine the conjecture that
the features of loose packed dominance and insertion graph dominance
found for large $D$ and fixed $k$ holds also for fixed $D$ and large $k$.

\subsection{Method}

The Monte-Carlo procedure used for this study is the same as that used by Ree
and Hoover (e.g. in \cite{ree1964a}). Random configurations are
generated by placing one point at the origin of some arbitrary
coordinate system, then the next is randomly placed within a
$D$-dimensional unit sphere centered on the first point, and so
on. The final configuration is then tested to see if it satisfies the
requirements imposed on distances by the diagram being calculated. For
each diagram we used batches of $10^6$ configurations, with enough
batches to give accuracy better than two percent up to a maximum of
$10000$ batches, and errors were calculated using \eqn{montecarloerr}.

\subsection{Results}

The evaluations of the following diagrams are presented in Tables
\ref{cstartab}--\ref{pinwheeltab}
of Appendix \ref{highord}
for values of $k$ up to 17 and values of $D$ up to 7;
$B_k[0,1], B_k[4,1], B_k[5,1], {\mathsf R}[k], {\mathsf R}_{4,2}[k], {\mathsf
R}_{4,1}[k], {\mathsf R}_{5,5}[k]$, and the ``pinwheel diagram'' 
in $B_k[k-1,1]$ obtained by
adding one point which is connected by $f$ bonds to all points in the
Ree-Hoover ring ${\mathsf R}[k-1].$ We note that for $k=5$ the pinwheel is
$B_5[4,1]$ and for $k=6$ it is $B_6[5,2].$

To investigate the conjecture of loose packed dominance for fixed $D$
 and large $k$ we use the tables of Appendix \ref{highord} to find for each $k$ and
 $D$ which diagram of this set is the largest. This is shown in
 \tab{largediagtab}. We also use Tables
  \ref{cstartab} and \ref{ringtab}
to compute the ratio ${\mathsf R}/B_k[0,1].$, and display the results in
\tab{ring_star}. 
From this table we see that for $D\geq 3$
 the Ree-Hoover ring ${\mathsf R}$ quickly dominates the complete star
 diagram as $k$ increases. However, for $D=2,$ even though the ratios
 are monotonically increasing for $k\geq 6$ the ring has not dominated
 the complete star at $k=17.$
 If we extrapolate the
increasing ratios in \tab{ring_star} for $D=2$ 
we can estimate that the ring will be larger for
approximately $k \sim 22$. 

\renewcommand{\arraystretch}{1.4}
\begin{table}[!htb]
\caption{Largest diagrams}
\label{largediagtab}
\begin{center}
\begin{tabular}{|r|c|c|c|c|c|c|c|}
\hline
&\multicolumn{7}{c|}{Dimension} \\
\hline
$k$ & 1 & 2 & 3 & 4 & 5 & 6 & $\infty$ \\
\hline
4 & \wigglya & \wigglya & \wigglya& \wigglya& \wigglya& \wigglya& ${\mathsf R}$ \\
5 & \wigglya & \wigglya & \wigglya & \wigglya & \wigglya & ${\mathsf R}$ & ${\mathsf R}$ \\
6 & \wigglya & \wigglya &\wigglya &\wigglya & ${\mathsf R}$ & ${\mathsf R}$ & ${\mathsf R}$ \\
7 & \wigglya & \wigglya &\wigglya & ${\mathsf R}$ & ${\mathsf R}$ & ${\mathsf R}$ & ${\mathsf R}$ \\
8 & \wigglya & \wigglya & ${\mathsf R \left(\insivb\right)}$ & ${\mathsf R}$ & ${\mathsf R}$ & ${\mathsf R}$ & ${\mathsf R}$ \\
9 & \wigglya & \wigglya &${\mathsf R \left(\insivb\right)}$ & ${\mathsf R \left(\insivb\right)}$ & ${\mathsf R}$ & ${\mathsf R}$ & ${\mathsf R}$ \\
10 & \wigglya & \wigglya &${\mathsf R \left(\insivb\right)}$ & ${\mathsf R \left(\insivb\right)}$ & ${\mathsf R}$ & & ${\mathsf R}$ \\
11 & \wigglya & \wigglya &${\mathsf R \left(\insivb\right)}$ & ${\mathsf R \left(\insivb\right)}$ & ${\mathsf R}$ & & ${\mathsf R}$ \\
12 & \wigglya & \wigglya &${\mathsf R \left(\insivb\right)}$ & ${\mathsf R \left(\insivb\right)}$ & ${\mathsf R \left(\insivb\right)}$ & & ${\mathsf R}$ \\
13 & \wigglya & \wigglya&${\mathsf R \left(\insivb\right)}$ & ${\mathsf R \left(\insivb\right)}$ & ${\mathsf R \left(\insivb\right)}$ & & ${\mathsf R}$ \\
14 & \wigglya & &${\mathsf R \left(\insivb\right)}$ & ${\mathsf R \left(\insivb\right)}$ & ${\mathsf R \left(\insivb\right)}$ & & ${\mathsf R}$ \\
15 & \wigglya & &${\mathsf R \left(\insivb\right)}$ & ${\mathsf R \left(\insivb\right)}$ & ${\mathsf R \left(\insivb\right)}$ & & ${\mathsf R}$ \\
\hline
\end{tabular}
\end{center}
\end{table}
\renewcommand{\arraystretch}{1}

\begin{table}[!htb]
\caption[]{${\mathsf{R}}/\wigglya$}
\label{ring_star}
\scriptsize
\begin{center}
\begin{tabular}{|c|c|c|c|c|c|}
\hline $k$ & \multicolumn{1}{c|}{$D = 2$} & \multicolumn{1}{c|}{$D = 3$} & \multicolumn{1}{c|}{$D = 4$} & \multicolumn{1}{c|}{$D = 5$} & \multicolumn{1}{c|}{$D = 6$} \\
\hline 4 & $-0.0299(2)$ & $-0.0942(4)$ & $-0.1964(8)$ &
$-0.3405(7)$ & $-0.535(2)$ \\
 5 & $\;\;\>0.0238(2)$ &
$\;\;\>0.1143(7)$ & $\;\;\>0.362(3)$ & $\;\;\>0.911(6)$ &
$\;\;\>1.97(1)$ \\
 6 & $-0.0230(4)$ & $-0.184(2)$ & $-0.91(1)$
& $-3.47(5)$ & $-11.7(2)$ \\
 7 & $\;\;\>0.0237(5)$ &
$\;\;\>0.337(7)$ & $\;\;\>2.92(6)$ & $\;\;\>18.8(5)$
& $\;\;\>85(2)$ \\
 8 & $-0.0277(6)$ & $-0.73(2)$ &
$-10.4(3)$ & $-100(3)$ &
$-780(30)$ \\
 9 & $\;\;\>0.0350(8)$ & $\;\;\>1.67(5)$
& $\;\;\>44(1)$ & $\;\;\>710(30)$ &
$\;\;\>11000(2000)$ \\
 10 & $-0.043(1)$ & $-4.1(1)$ &
$-177(7)$ & $-5100(800)$ & \\
 11 &
$\;\;\>0.056(2)$ & $\;\;\>10.3(3)$ &
$\;\;\>860(80)$ & & \\
 12 & $-0.076(2)$ &
$-26.0(7)$ & $-3000(800)$ & & \\
 13 &
$\;\;\>0.102(3)$ & $\;\;\>60(5)$ & & & \\
 14 &
$-0.133(5)$ & $-160(30)$ & & & \\
 15 &
$\;\;\>0.18(1)$ & & & & \\
 16 & $-0.24(2)$ & & & & \\
 17 &
$\;\;\>0.39(5)$ & & & & \\ 
\hline
\end{tabular}
\end{center}
\normalsize
\end{table}

Further information is obtained from the ratios of $B_k[4,1]/B_k[0,1]$ and
$B_k[5,1]/B_k[0,1]$ plotted in Figs. \ref{rh2graph} and \ref{rh3graph} of Appendix \ref{highordgraphs}.
In these figures it is seen that the ratios $B_k[4,1]/B_k[0,1]$ and 
$B_k[5,1]/B_k[0,1]$ increase linearly with $k.$ Thus while for the
values of $k$ studied the ratios only rarely exceed one it appears
that if we extrapolate to sufficiently large $k$ that the inequalities in
\eqn{lpdom} will be satisfied for $m,m'=0,4,5.$

The relative size of diagrams in the class $B_k[k,i]$ and $B_k[k-1,i]$
is studied by comparing the Ree-Hoover ring diagram ${\mathsf R}[k]$ of \tab{ringtab}
with the corresponding pinwheel diagram of \tab{pinwheeltab}. It
is clear that in each dimension that for sufficiently large $k$ the pinwheel diagram
will vanish, and although this can't be shown numerically one can see
the ratio of the pinwheel to the ring ${\mathsf R}[k]$ decreases rapidly
as $k$ increases.

We conclude this discussion of results by examining
the relative magnitudes of insertion diagrams in $B_k[k,i].$
In Figs. \ref{loose4bgraph}, \ref{loose4agraph}, and
\ref{loose5egraph} of Appendix \ref{highordgraphs} we plot the ratios
${\mathsf R}_{4,2}/{\mathsf R}$,
${\mathsf R}_{4,1}/{\mathsf R}$,
and ${\mathsf R}_{5,5}/{\mathsf R}$ as a function of $k$ for various $D.$
We see that for sufficiently large $k$ all three ratios have a
linear increase with respect to $k$.
We also note that even though the ratio ${\mathsf R}_{4,1}/{\mathsf R}_{4,2}$
is small in low dimensions, it approaches a non-zero constant as
$k\rightarrow \infty$, and for sufficiently large $D$ the ratio
is in fact greater that unity.

The examination of all available data supports the
following conjectures
\begin{conjecture}
The loose packed diagrams in $B_k[k,i]$ dominate all other diagrams
for large $k$ when $D\geq 3$.
\label{loosepackeddominance}
\end{conjecture}
In connection with this conjecture we note that 
the three largest diagrams for $B_6$ in $D=5,6$ are ${\mathsf R}_{5,5}$, ${\mathsf R}_{4,1}$, and ${\mathsf
R}$ and that the sum of these three largest diagrams shares
the property with $B_6$ itself of changing sign between $D=5$ and
$D=6.$ We restrict the conjecture to dimensions $D \ge 3$ because at
this stage there is not yet enough data to support the case for $D=2$.
\begin{conjecture}
The insertion diagrams all have the same
exponential rate of growth as the Ree-Hoover ring. 
\end{conjecture}

\section{Estimates of the radius of convergence}
\label{radiisec}

The dominance of loose packed diagrams for fixed $D\geq
3$ and large $k$ may be used to discuss the question of 
sign change in the virial 
coefficients and the radius of convergence of the virial  expansion.
Of greatest importance is the relation of the radius of convergence
to the packing fractions $\eta_f$ at which freezing occurs which have been
numerically determined as $\eta_f=0.49$ in $D=3$ 
\cite{alder1960a,hoover1968a}, $\eta_f=0.31$ in $D=4$ 
and $\eta_f=0.19$ in $D=5$ as obtained from
\cite{michels1984a,finken2001a}.

It may be that there are sequences of loose packed diagrams that grow
faster than the ring, but for low order the largest diagrams can be
characterized as insertion diagrams and these appear to have the same
exponential rate of growth as the ring. Hence we will concentrate on the Ree-Hoover ring
${\mathsf R}$. We first note that the absolute value of this diagram must be
strictly less than the absolute value of the corresponding Mayer
diagram where all $\tilde f$ bonds are replaced by unity. The
resulting Mayer ring diagram has been long ago evaluated 
\cite{katsura1963a}
in terms of
Bessel functions of the first kind $J_{D/2}(x)$ and thus we obtain
\be
(-1)^{k-1}{\mathsf R}[k]/B_2^{k-1}\leq{\frac{(k-1)(2\pi)^{kD/2}}{ 2k\Omega_{D-1}^{k-2}}}
\int_0^\infty dx \; x^{D-1}\left[\frac{J_{D/2}(x)}{x^{D/2}}\right]^{k}
\ee
 The large
$k$ behavior of this integral is easily obtained by steepest descents
by noting that the maximum value of $J_{D/2}(x)/x^{D/2}$ occurs at $x=0.$
Thus we have as $k\rightarrow \infty$
\be
(-1)^{k-1}{\mathsf R}[k]/B_2^{k-1}\leq{\frac{(k-1)(1+D/2)^{D/2}}{ k^{1+D/2}\Gamma(1+D/2)}}
\> 2^{k-2}
\ee
and hence ${\mathsf R}[k]$ satisfies the bound 
\be
|{\mathsf R}[k+1]/(B_2{\mathsf R}[k])|\leq 2.
\label{rcbound}
\ee
This leads to a packing fraction at the radius of convergence 
of the sum of Ree-Hoover diagrams  of $\eta_{rh}=2^{-D}$
  which is substantially greater than the lower bound $0.145/2^D$ of 
Lebowitz and Penrose~\cite{lebowitz1964a}
but which is still much smaller than the freezing densities $\eta_f.$
This value of $\eta_{rh}$ is of the same order of magnitude
as the lower bounds on the termination density due to Baram and Fixman~\cite{baram1993a}
which are $\eta_t > 0.25838$ for $D=2$ and $\eta_t > 0.12681$ for $D=3$.This is compatible with the pressure being analytic for positive
values of $\eta$ less than $\eta_f$ because the alternations of sign
of ${\mathsf R}[k]$ puts the leading singularity on the negative $\eta$ axis.

However the ratios $\left|{\mathsf R}[k+1]/(B_2{\mathsf R}[k])\right|$
obtained from \tab{ringtab} are substantially below the bound
\eqn{rcbound}. In fact 
we see from 
each $D\geq 4$ that there is some value of $k$ such that for all
greater values of $k$ the values of ${\mathsf R}/B_2^{k-1}$ increase. The
order $k$ must surely be greater than this value before it can be
claimed that the asymptotic regions has been achieved. 
If we use that maximum ratios as obtained from \tab{ringtab}
we estimate that for $D=4$ we have $\eta_{rh}\sim 0.12$ and for $D=5$
we have $\eta_{rh}\sim 0.052.$  In $D=3$ if we assume that the data of
\tab{ringtab} extrapolates to a constant as $k\rightarrow \infty$
then the radius of convergence of the Ree-Hoover ring is $\eta_{rh}=0.25$.
At most the ratios are bounded below by $0.91$ which leads to $\eta_{rh}=0.27$.
All of these estimated radii of convergence are substantially 
less than the freezing densities $\eta_f.$ 

In order for the radius of convergence of the virial series to be
larger than these estimates obtained from the Ree-Hoover ring
there must be cancellations between diagrams in the class $B_k[k,i].$
Such cancellations can occur because, for example, the diagrams
${\mathsf R}_{4,2}[k]$ and ${\mathsf R}_{5,5}[k]$ which have magnitudes comparable to
${\mathsf R}[k]$ have signs opposite to ${\mathsf R}[k].$ Furthermore from
\tab{virialtable} we see that $B_k$ is indeed less in magnitude than ${\mathsf
R}[k]$ for $k=6,7,8.$  What cannot be inferred from the existing data
is whether or not when $k$ is sufficiently large that the diagrams are
in their asymptotic region that the cancellation is severe enough to
reduce the exponential rate of growth of the ratios $B_{k+1}/B_{k}.$ 
Unless the cancellation becomes sufficiently great for the radius of
convergence to be larger than the freezing density then the leading
singularity cannot be on the real $\eta$ axis and there must
be oscillation in the signs of the virial coefficients.

\section{Approximate equations of state for hard spheres}
\label{eqstatesec}

For over 40 years the eight virial coefficients of \tab{virialtable} have been
used to inspire many approximate equations of state for the low
density phase of hard spheres. These approximates may be grouped 
by the location of their leading singularity  
into the following three classes:

\vspace{0.2cm}

\noindent {\bf 1: High order poles at $\eta=1$}

Examples of these are the proposals of Thiele~\cite{thiele1963a}
\be
Pv/k_BT={\frac{1+2\eta+3\eta^2}{ (1-\eta)^2}},
\ee
Reiss, Frisch, and Lebowitz~\cite{reiss1959a}, and Wertheim~\cite{wertheim1963a}
\be
Pv/k_BT={\frac{1+\eta+\eta^2}{(1-\eta)^3}},
\ee
Guggenheim~\cite{guggenheim1965a}
\be
Pv/k_BT={\frac{1}{(1-\eta)^4}},
\ee
and the proposal of Carnahan and Starling~\cite{carnahan1969a} 
\be
Pv/k_BT=\frac{1+\eta+\eta^2-\eta^3}{(1-\eta)^3}.
\label{carnstarl}
\ee
We do note that Torquato \cite{torquato1995a,torquato1995b} proposes an equation of state which agrees with \eqn{carnstarl} for $\eta < \eta_f$ but which is of a different form for $\eta > \eta_f$.

\vspace{0.2cm}

\noindent {\bf 2: Simple poles at}~\cite{goldman1988a,hoste1984a} 
{\bf or near the packing fraction} $\eta_{cp}=0.74048\cdots$ {\bf of closest packed hard spheres}~\cite{ree1964a,hoover1968a,vanrensburg1993a,wang2002a}

In particular the  Pad{\'e} analysis made in 
\cite{vanrensburg1993a} has simple poles at
\be
\eta=(1.22\pm0.09i)\eta_{cp}
\ee
which leads to sign oscillations beginning with $B_{45}$.

\vspace{0.2cm}

{\bf 3: A fractional power law divergence at or near the ``random
close packed'' density}  $\eta_{{rcp}}=0.64$ {\bf as defined by}~\cite{bernal1960a,bernal1964a,scott1960a,finney1970a}

These approximates are obtained from a D-log Pad{\'e} analysis and are 
(generalizations) of the form
\be
Pv/k_BT=A(\eta-\eta_{{rcp}})^{-s}
\ee
As an example $s$ is estimated as 1 in \cite{lefevre1972a} as $0.678$ in
\cite{ma1986a}
and $0.76$ in \cite{song1988a}. In \cite{jasty1987a} other values 
of $\eta_{{rcp}}$ are chosen and the values of $s$ lie in 
the range $0.6\leq s\leq 0.9$
depending on the approximation used.

All these approximate equations of state share the feature that their
leading singularity is at a value of $\eta$ which is greater than the
freezing density $\eta_f=0.49$ and indeed is even greater than the solid
end of the phase transition $\eta_s=0.542.$ This fact has lead to the
assumption that the virial expansion is analytic at the freezing
transition $\eta_f$ and this qualitative feature is incorporated into
most of the phenomenological theories used to describe freezing  
\cite{kirkwood1940a,
ramakrishnan1979a,haymet1986a,laird1987a}. It is therefore of 
great importance that the estimates of the radius of convergence made
above on the basis of loose packed dominance and the assumption of no
cancellation estimated the radius of convergence at no more than $0.27$.  

The estimate of the radius of convergence relies on values of $k$
beyond the first eight virial coefficients used to obtain the
approximate equations of state. Therefore it is fair to say
that none of the approximates incorporates the true large $k$ behavior
of the virial coefficients. Thus even if there is cancellation for
large $k$ for the set of close packed diagrams none of the
approximates is based on computations which can observe these
cancellations and hence no known approximate equation of state can
be considered reliable. We therefore conclude that at present there
exists no evidence to support the claim that the virial expansion has
a radius of convergence greater than the freezing density $\eta_f.$ 

\clearpage

\appendix

\section{Individual diagram contributions for $B_4$, $B_5$, and $B_6.$}
\label{inddiags}

In this appendix we tabulate the contributions of the individual
Ree-Hoover diagrams to the virial coefficients for $B_4$, $B_5$, and $B_6.$

\begin{table}[!htbp]
\caption{Individual diagram contributions to $B_4$.}
\label{b4diagtab}
\vspace{0.2cm}
\footnotesize
\begin{center}
\begin{tabular}{|r|l|l|l|}
\hline
$D$ & $B_4[0,1]/B_2^3$ & $B_4[4,1]B_2^3$ & $\;\;\>B_4/B_2^3$ \\
\hline
 3 & $  0.31673(2)$ &  $  -0.029781(8)$ & $\;\;\>0.2869495\cdots$ \\
 4 & $  0.1888655\cdots$ & $  -0.0370195\cdots$ & $\;\;\>0.1518460\cdots$ 
\\
 5 & $  0.115211(3)$ & $  -0.039233(3)$ & $\;\;\>0.075978(4)$ \\
 6 & $  0.0714700\cdots$ & $  -0.0381069\cdots$ & $\;\;\>0.03336314\cdots$ 
\\
 7 & $  0.044927(2)$ & $  -0.035055(3)$ & $\;\;\>0.009873(3)$ \\
 7.7 & $0.032669(2)$ & $  -0.032331(3)$ & $\;\;\>0.000338(3)$ \\
 7.8 & $0.031227(2)$ & $  -0.031920(3)$ & $ -0.000693(3)$ \\
 8 & $  0.0285344\cdots$ & $  -0.0310921\cdots$ & $ -0.0025576\cdots$ \\
 9 & $  0.018286(1)$ & $  -0.026861(3)$ & $ -0.008575(3)$ \\
 10 & $ 0.0117986\cdots$ & $  -0.0227611\cdots$ & $ -0.0109624\cdots$ \\
 11 & $ 0.0076638(8)$ & $ -0.018997(3)$ & $ -0.011333(3)$ \\
 12 & $ 0.0050018\cdots$ & $ -0.0156721\cdots$ & $ -0.0109624\cdots$ \\
 13 & $ 0.0032819(5)$ & $ -0.012805(2)$ & $ -0.009523(2)$ \\
 14 & $ 0.0021615(4)$ & $ -0.010381(2)$ & $ -0.008220(2)$ \\
 15 & $ 0.0014288(3)$ & $ -0.008362(2)$ & $ -0.006933(2)$ \\
 20 & $0.00018830(6)$ & $-0.0026504(7)$ & $-0.0024621(7)$ \\
 25 & $0.00002615(1)$ & $-0.0007841(3)$ & $-0.0007580(3)$ \\
 30 & $3.763(3)\times10^{-6}$ & $-0.0002233(1)$ & $-0.0002196(1)$ \\
 35 & $5.560(7)\times10^{-7}$ & $-0.00006217(3)$ & $-0.00006162(3)$ \\
 40 & $8.38(1)\times10^{-8}$ & $-0.00001705(1)$ & $-0.00001697(1)$ \\
 45 & $1.284(3)\times10^{-8}$ & $-4.631(4)\times 10^{-6}$ & 
$-4.618(4)\times 10^{-6}$ \\
 50 & $1.992(5)\times10^{-9}$ & $-1.249(1)\times 10^{-6}$ & 
$-1.247(1)\times 10^{-6}$ \\
\hline
\end{tabular}
\end{center}
\normalsize
\end{table}

\begin{table}[!ht]
\caption{Individual diagram contributions to $B_5$. Values for discs
and spheres taken from \cite{ree1964a}. The contributions
from the Ree-Hoover ring diagram $B_5[5,2]$ are underlined.}
\label{b5diagtab}
\vspace{0.2cm} \footnotesize
\begin{center}
\begin{tabular}{|r|l|l|l|l|}\hline
&discs&spheres&$D=4$&$D=5$\\ \hline
$B_5/B_2^4$&$\;\;\>0.3336$&$\;\;\>0.1103$&$\;\;\>0.03565(5)$ &
$\;\;\>0.01297(1)$\\
\hline 
$B_5[0,1]/B_2^4$&$\;\;\>0.3618$&$\;\;\>0.1422$&$\;\;\>0.059015(9)$&$\;\;\>0.025442(1)$\\
\hline
$B_5[4,1]/B_2^4$&$-0.0266$&$-0.0314$&$-0.02650(2)$&$-0.019184(5)$\\
\hline 
$B_5[5,1]/B_2^4$&$-0.0102$&$-0.0165$&$-0.01762(4)$&$-0.015511(4)$\\
$B_5[5,2]/B_2^4$&$\;\;\>\underline{0.0086}$&$\;\;\>\underline{0.0162}$
&$\;\;\>\underline{0.02131(2)}$&$\;\;\>\underline{0.022980(7)}$
\\
$B_5[5,3]/B_2^4$&$\;\;\>0$&$-0.0002$&$-0.0005498(5)$&$-0.0007622(3)$\\
\hline
\hline
&$D=6$&$D=7$&$D=8$&$D=50$\\ \hline
$B_5/B_2^4$& $\;\;\>0.007528(8)$&$\;\;\>0.007071(7)$&
$\;\;\>0.007429(6)$ & $\;\;\>2.17(1)\times 10^{-8}$\\
\hline 
$B_5[0,1]/B_2^4$&$\;\;\>0.0112852(7)$&$\;\;\>0.0051189(4)$&$\;\;\>0.0023640(3)$& 
$\;\;\>1.67(8)\times 10^{-15}$\\
\hline
$B_5[4,1]/B_2^4$&$-0.012899(4)$&$-0.008296(3)$&$-0.005185(2)$&$-6.1(1)\times 10^{-13}$\\
\hline 
$B_5[5,1]/B_2^4$&$-0.012351(3)$&$-0.009220(3)$&$-0.006588(3)$& 
$-1.38(3)\times 10^{-11}$\\
$B_5[5,2]/B_2^4$
&$\;\;\>\underline{0.022332(6)}$&$\;\;\>\underline{0.020277(6)}$&$\;\;\>\underline{0.017522(6)}$& 
$\;\;\>\underline{2.17(1)\times 10^{-8}}$\\
$B_5[5,3]/B_2^4$&$-0.0008395(4)$&$-0.0008090(4)$&$-0.0007149(4)$& 
$-4.84(7)\times 10^{-11}$\\
\hline
\end{tabular}
\end{center}
\normalsize
\end{table}

\begin{table}[!ht]
\caption{Individual diagram contributions to $B_6$. Values for discs
and spheres taken from \cite{ree1964a}.
The contributions from the Ree-Hoover ring diagram $B_6[6,3]$ are
underlined. For $D=50$ when the value $x.xx \times 10^m$ has $m \le -16$ we
write $\sim 10^{m}$.}
\label{b6diagtab}
\vspace{0.2cm} 
\scriptsize%
\begin{center}
\begin{tabular}{|r|l|l|l|l|}\hline
&discs&spheres&$D=5$&$D=6$\\ \hline
$B_6/B_2^5$&$\;\;\>0.1994$&$\;\;\>0.0386$&$\;\;\>0.00102(8)$&$-0.00176(2)$\\
\hline 
$B_6[0,1]/B_2^5$&$\;\;\>0.2292$&$\;\;\>0.0588$&$\;\;\>0.0048248(9)$&$\;\;\>0.0014771(1)$\\
\hline
$B_6[4,1]/B_2^5$&$-0.0273$&$-0.0212$&$-0.00569(2)$&$-0.002600(3)$\\
\hline 
$B_6[5,1]/B_2^5$&$-0.0191$&$-0.0187$&$-0.00719(1)$&$-0.003800(4)$\\
$B_6[5,2]/B_2^5$&$\;\;\>0.0090$&$\;\;\>0.0099$&$\;\;\>0.00498(2)$&$\;\;\>0.003038(4)$\\
$B_6[5,3]/B_2^5$&$\;\;\>0$&$-0.0002$&$-0.000244(1)$&$-0.0001759(3)$\\
\hline 
$B_6[6,1]/B_2^5$&$\;\;\>0.0088$&$\;\;\>0.0132$&$\;\;\>0.01029(4)$&$\;\;\>0.00735(1)$\\
$B_6[6,2]/B_2^5$&$\;\;\>0.0077$&$\;\;\>0.0121$&$\;\;\>0.01064(4)$&$\;\;\>0.008204(8)$\\
$B_6[6,3]/B_2^5$&$\underline{-0.0051}$&$-\underline{0.0109}$
&$\underline{-0.01702(5)}$&$\underline{-0.01693(2)}$
\\
$B_6[6,4]/B_2^5$&$-0.0019$&$-0.0027$&$-0.001520(4)$&$-0.0009332(8)$\\
$B_6[6,5]/B_2^5$&$-0.0010$&$-0.0022$&$-0.001559(9)$&$-0.001001(2)$\\
$B_6[6,6]/B_2^5$&$-0.0009$&$-0.0011$
&$-0.000553(1)$&$-0.0003335(3)$\\
$B_6[6,7]/B_2^5$&$-0.0005$&$-0.0007$&$-0.000437(1)$&$-0.0002948(3)$\\
$B_6[6,8]/B_2^5$&$\;\;\>0.0004$&$\;\;\>0.0008$&$\;\;\>0.000712(2)$&$\;\;\>0.0004958(8)$\\
$B_6[6,9]/B_2^5$&$\;\;\>0.0001$&$\;\;\>0.0012$&$\;\;\>0.00256(2)$&$\;\;\>0.002356(7)$\\
$B_6[6,10]/B_2^5$&$\;\;\>0.0000$&$\;\;\>0.0002$&$\;\;\>0.0003086(8)$&$\;\;\>0.0002481(2)$\\
$B_6[6,11]/B_2^5$&$-0.0000$&$-0.0003$&$-0.0002596(7)$&$-0.0001587(1)$\\
$B_6[6,12]/B_2^5$&$\;\;\>0$&$-0.0002$&$-0.000472(2)$&$-0.0003931(6)$\\
$B_6[6,13]/B_2^5$&$\;\;\>0$&$\;\;\>0.0002$&$\;\;\>0.000318(4)$&$\;\;\>0.000285(1)$\\
$B_6[6,14]/B_2^5$&$\;\;\>0$&$-0.0000$&$-0.0000991(5)$&$-0.0001007(2)$\\
$B_6[6,15]/B_2^5$&$\;\;\>0$&$\;\;\>0.0000$&$\;\;\>0.000046(2)$&$\;\;\>0.0000486(9)$\\
$B_6[6,16]/B_2^5$&$\;\;\>0$&$\;\;\>0.0004$&$\;\;\>0.00138(1)$&$\;\;\>0.001463(2)$\\
$B_6[6,17]/B_2^5$&$\;\;\>0$&$-0?$&$-2.00(1)\times
10^{-6}$&$-3.176(6)\times 10^{-6}$\\
$B_6[6,18]/B_2^5$&$\;\;\>0$&$\;\;\>0$&$\;\;\>2.59(1)\times
10^{-7}$&$\;\;\>3.696(8)\times 10^{-7}$\\
\hline
\end{tabular}

\vspace{0.2cm}

\begin{tabular}{|r|l|l|l|}\hline
&$D=7$&$D=8$&$D=50$\\ \hline
$B_6/B_2^5$&$-0.00352(2)$&$-0.00451(1)$&$-7.6(2)\times 
10^{-10}$\\
\hline 
$B_6[0,1]/B_2^5$&$\;\;\>0.00046725(4)$&$\;\;\>0.00015174(2)$& 
$\sim 10^{-22}$\\
\hline
$B_6[4,1]/B_2^5$&$-0.001145(1)$&$-0.0004946(6)$&$\sim 
10^{-20}$\\
\hline 
$B_6[5,1]/B_2^5$&$-0.001902(2)$&$-0.0009209(9)$&$\sim 
10^{-18}$\\
$B_6[5,2]/B_2^5$&$\;\;\>0.001752(2)$&$\;\;\>0.0009748(9)$&$\sim 
10^{-16}$\\
$B_6[5,3]/B_2^5$&$-0.0001124(2)$&$-0.0000665(1)$&$\sim 
10^{-18}$\\
\hline 
$B_6[6,1]/B_2^5$&$\;\;\>0.004866(6)$&$\;\;\>0.003063(3)$&$\;\;\>6.7(7)\times 
10^{-15}$\\
$B_6[6,2]/B_2^5$&$\;\;\>0.005887(5)$&$\;\;\>0.004019(3)$&$\;\;\>1.6(1)\times 
10^{-13}$\\
$B_6[6,3]/B_2^5$&$\underline{-0.01540(2)}$&$\underline{-0.01318(2)}$ 
&$\underline{-7.6(2)\times 10^{-10}}$\\
$B_6[6,4]/B_2^5$&$-0.0005349(5)$&$-0.0002926(3)$&$\sim 
10^{-18}$\\
$B_6[6,5]/B_2^5$&$-0.000589(1)$&$-0.0003277(6)$&$\sim 
10^{-18}$\\
$B_6[6,6]/B_2^5$&$-0.0001883(1)$&$-0.00010175(8)$&$\sim 10^{-18}$\\
$B_6[6,7]/B_2^5$&$-0.0001858(2)$&$-0.0001114(2)$&$\sim 
10^{-17}$\\
$B_6[6,8]/B_2^5$&$\;\;\>0.0003158(4)$&$\;\;\>0.0001895(2)$&$\sim 
10^{-17}$\\
$B_6[6,9]/B_2^5$&$\;\;\>0.001912(4)$&$\;\;\>0.001428(3)$&$\;\;\>5.3(5)\times 
10^{-14}$\\
$B_6[6,10]/B_2^5$&$\;\;\>0.0001778(1)$&$\;\;\>0.0001180(1)$&$\sim 
10^{-16}$\\
$B_6[6,11]/B_2^5$&$-0.00008782(7)$&$-0.00004567(4)$&$\sim 
10^{-20}$\\
$B_6[6,12]/B_2^5$&$-0.0002848(4)$&$-0.0001889(2)$&$\sim 
10^{-17}$\\
$B_6[6,13]/B_2^5$&$\;\;\>0.0002231(6)$&$\;\;\>0.0001595(4)$&$\sim 
10^{-16}$\\
$B_6[6,14]/B_2^5$&$-0.0000868(2)$&$-0.0000671(1)$&$\sim 
10^{-16}$\\
$B_6[6,15]/B_2^5$&$\;\;\>0.0000447(4)$&$\;\;\>0.0000371(3)$&$\;\;\>2.2(3)\times 
10^{-15}$\\
$B_6[6,16]/B_2^5$&$\;\;\>0.001358(2)$&$\;\;\>0.001147(2)$&$\;\;\>1.34(8)\times 
10^{-12}$\\
$B_6[6,17]/B_2^5$&$-3.774(8)\times 
10^{-6}$&$-3.758(9)\times 10^{-6}$&$-1.5(1)\times 10^{-15}$\\
$B_6[6,18]/B_2^5$&$\;\;\>3.944(9)\times 
10^{-7}$&$\;\;\>3.524(9)\times 10^{-7}$&$\sim 10^{-18}$\\
\hline
\end{tabular}
\end{center}
\normalsize
\end{table}

\clearpage

\section{Numerical values of selected diagrams to high order.}
\label{highord}

In this appendix we tabulate the results of Monte-Carlo evaluations of
selected diagrams $B_k[m,i]$ to orders up to $k=17$.

\begin{table}[!htbp]
\caption[]{$\wigglya/B_2^{k-1}=B_k[0,1]/B_2^{k-1}$}
\label{cstartab}
\vspace{0.2cm}
\footnotesize
\begin{center}
\begin{tabular}{|l|l|l|l|l|l|}
\hline
$k$ & \multicolumn{1}{c|}{$D = 2$} & \multicolumn{1}{c|}{$D = 3$} & \multicolumn{1}{c|}{$D = 4$} & \multicolumn{1}{c|}{$D = 5$} & \multicolumn{1}{c|}{$D = 6$} \\
\hline
3 & $0.7821(1)$ & $0.6248(2)$ & $0.5063(2)$ & $0.4143(2)$ & $0.3410(2)$ \\
4 & $0.5488(4)$ & $0.3166(3)$ & $0.1888(2)$ & $0.1153(2)$ & $0.0713(2)$ \\
5 & $0.3620(3)$ & $0.1420(2)$ & $0.0591(2)$ & $0.02522(8)$ & $0.01121(7)$ \\
6 & $0.2292(3)$ & $0.0593(2)$ & $0.01648(6)$ & $0.00487(6)$ & $0.00148(2)$ \\
7 & $0.1412(3)$ & $0.0233(2)$ & $0.00424(6)$ & $0.00076(1)$ & $0.000170(3)$ \\
8 & $0.0844(4)$ & $0.0087(2)$ & $0.00101(2)$ & $0.000129(3)$ & $1.81(5)\times10^{-5}$ \\
9 & $0.0505(4)$ & $0.00315(6)$ & $0.000226(5)$ & $1.78(7)\times10^{-5}$ & $1.3(2)\times10^{-6}$ \\
10 & $0.0293(4)$ & $0.00111(2)$ & $5.2(2)\times10^{-5}$ & $2.5(4)\times10^{-6}$ & \\
11 & $0.0170(3)$ & $0.000380(8)$ & $1.0(1)\times10^{-5}$ & & \\
12 & $0.0097(2)$ & $0.000128(3)$ & $2.7(7)\times10^{-6}$ & & \\
13 & $0.0053(1)$ & $5.2(4)\times10^{-5}$ & & & \\
14 & $0.00304(6)$ & $1.7(3)\times10^{-5}$ & & & \\
15 & $0.00179(4)$ & & & & \\
16 & $0.00098(2)$ & & & & \\
17 & $0.00055(1)$ & & & & \\
\hline
\end{tabular}
\end{center}
\normalsize
\end{table}
 
\begin{table}[!htbp]
\caption[]{$\wigglyb/B_2^{k-1}=B_k[4,1]/B_2^{k-1}$}
\label{rh2tab}
\vspace{0.2cm}
\footnotesize
\begin{center}
\begin{tabular}{|l|l|l|l|l|l|}
\hline
$k$ & \multicolumn{1}{c|}{$D = 2$} & \multicolumn{1}{c|}{$D = 3$} & \multicolumn{1}{c|}{$D = 4$} & \multicolumn{1}{c|}{$D = 5$} & \multicolumn{1}{c|}{$D = 6$} \\
\hline
4 & $-0.01644(5)$ & $-0.02981(9)$ & $-0.0370(1)$ & $-0.0391(1)$ & $-0.0382(1)$ \\
5 & $-0.0264(3)$ & $-0.0316(2)$ & $-0.0270(2)$ & $-0.0189(2)$ & $-0.0130(2)$ \\
6 & $-0.0285(5)$ & $-0.0219(4)$ & $-0.0117(2)$ & $-0.0059(1)$ & $-0.00256(5)$ \\
7 & $-0.0239(5)$ & $-0.0114(2)$ & $-0.00403(8)$ & $-0.00130(3)$ & $-0.00039(1)$ \\
8 & $-0.0183(4)$ & $-0.0056(1)$ & $-0.00120(5)$ & & \\
9 & $-0.0123(3)$ & $-0.0025(1)$ & & & \\
10 & $-0.0086(4)$ & & & & \\
11 & $-0.0056(2)$ & & & & \\
12 & $-0.0038(2)$ & & & & \\
13 & $-0.0023(3)$ & & & & \\
\hline
\end{tabular}
\end{center}
\normalsize
\end{table}
 
\begin{table}[!htbp]
\caption[]{${\wigglyc}/B_2^{k-1}=B_k[5,1]/B_2^{k-1}$}
\label{rh3tab}
\vspace{0.2cm}
\footnotesize
\begin{center}
\begin{tabular}{|l|l|l|l|l|l|}
\hline
$k$ & \multicolumn{1}{c|}{$D = 2$} & \multicolumn{1}{c|}{$D = 3$} & \multicolumn{1}{c|}{$D = 4$} & \multicolumn{1}{c|}{$D = 5$} & \multicolumn{1}{c|}{$D = 6$} \\
\hline
5 & $-0.01016(6)$ & $-0.0164(1)$ & $-0.01741(9)$ & $-0.01550(9)$ & $-0.0123(1)$ \\
6 & $-0.0188(3)$ & $-0.0189(2)$ & $-0.0124(2)$ & $-0.0069(1)$ & $-0.00386(8)$ \\
7 & $-0.0227(3)$ & $-0.0130(3)$ & $-0.0053(1)$ & $-0.00200(4)$ & $-0.00070(1)$ \\
8 & $-0.0205(4)$ & $-0.0072(1)$ & $-0.00199(5)$ & & \\
9 & $-0.0163(3)$ & $-0.0036(1)$ & & & \\
10 & $-0.0115(5)$ & & & & \\
11 & $-0.0083(2)$ & & & & \\
12 & $-0.0056(3)$ & & & & \\
13 & $-0.0043(5)$ & & & & \\
\hline
\end{tabular}
\end{center}
\normalsize
\end{table}

\begin{table}[!htbp]
\caption[]{${\mathsf R}/B_2^{k-1}.$ The underline marks the approximate
location of the minimum value.} 
\label{ringtab}
\vspace{0.2cm}
\footnotesize
\begin{center}
\begin{tabular}{|l|l|l|l|}
\hline
$k$ & \multicolumn{1}{c|}{$D = 2$} & \multicolumn{1}{c|}{$D = 3$} & \multicolumn{1}{c|}{$D = 4$}  \\
\hline
3 & $\;\;\>0.7824(2)$ & $\;\;\>0.6248(2)$ & $\;\;\>0.5063(2)$ \\
4 & $-0.01639(9)$ & $-0.0298(1)$ & $-0.0371(1)$ \\
5 & $\;\;\>0.00860(6)$ & $\;\;\>0.01623(9)$ & $\;\;\>0.0214(2)$ \\
6 & $-0.00526(8)$ & $-0.0109(1)$ & $-0.0150(2)$ \\
7 & $\;\;\>0.00335(6)$ & $\;\;\>0.0078(2)$ & $\;\;\>0.0124(2)$ \\
8 & $-0.00234(5)$ & $-0.0064(1)$ & $-0.0106(2)$ \\
9 & $\;\;\>0.00177(4)$ & $\;\;\>0.0053(1)$ & $\;\;\>0.0098(2)$ \\
10 & $-0.00125(3)$ & $-0.00452(9)$ & $-0.0091(2)$\\
11 & $\;\;\>0.00095(2)$ & $\;\;\>0.00392(8)$ & $\;\;\>0.0089(2)$ \\
12 & $-0.00074(1)$ & $-0.00333(7)$ & $\underline{-0.0083(2)}$\\
13 & $\;\;\>0.00055(1)$ & $\;\;\>0.00313(8)$ & $\;\;\>0.0086(2)$ \\
14 & $-0.00041(1)$ & $-0.0027(1)$ & $-0.0086(2)$ \\
15 & $\;\;\>0.00033(2)$ & $\;\;\>\underline{0.0026(1)}$ & $\;\;\>0.0087(3)$\\
16 & $-0.00023(2)$ & & \\
17 & $\;\;\>0.00021(3)$ & &\\
\hline
\hline
$k$ & \multicolumn{1}{c|}{$D = 5$} & \multicolumn{1}{c|}{$D = 6$} & \multicolumn{1}{c|}{$D = 7$} \\
\hline
3 & $\;\;\>0.4139(2)$ & $\;\;\>0.3409(2)$ & $\;\;\>0.2822(2)$ \\
4 & $-0.03925(5)$ & $-0.03815(8)$ & $-0.0351(1)$ \\
5 & $\;\;\>0.0230(1)$ & $\;\;\>0.02210(6)$ & $\;\;\>0.02025(9)$ \\
6 & $-0.01689(9)$ & $-0.0173(1)$ & $-0.0153(2)$ \\
7 & $\;\;\>0.0142(3)$ & $\;\;\>0.0144(2)$ & $\;\;\>0.0135(2)$ \\
8 & $-0.0129(3)$ 
& $-0.0141(3)$ & $\underline{-0.0132(2)}$ \\
9 & $\;\;\>0.0126(2)$ & $\;\;\>\underline{0.0138(3)}$ & $\;\;\>0.0132(3)$ \\
10 & $\underline{-0.0126(3)}$ & $-0.0150(3)$ & $-0.0143(3)$ \\
11 & $\;\;\>0.0128(3)$ & $\;\;\>0.0162(3)$ & \\
12 & $-0.0134(3)$ & $-0.0172(3)$ & \\
13 & $\;\;\>0.0142(3)$ & $\;\;\>0.0201(4)$ & \\
14 & $-0.0166(3)$ & $-0.0238(5)$ & \\
15 & $\;\;\>0.0183(4)$ & $\;\;\>0.0281(6)$ & \\
\hline
\end{tabular}
\end{center}
\normalsize
\end{table}
 
\begin{table}[!htbp]
\caption[]{${\mathsf R \left(\insivb\right)}/B_2^{k-1}={\mathsf
R}_{4,2}/B_2^{k-1}.$ The underline marks the approximate location 
of the minimum value.}
\label{ivbtab}
\vspace{0.2cm}
\footnotesize
\begin{center}
\begin{tabular}{|l|l|l|l|l|l|}
\hline
$k$ & \multicolumn{1}{c|}{$D = 2$} & \multicolumn{1}{c|}{$D = 3$} & \multicolumn{1}{c|}{$D = 4$} & \multicolumn{1}{c|}{$D = 5$} & \multicolumn{1}{c|}{$D = 6$} \\
\hline
5 & $-0.01017(4)$ & $-0.01654(7)$ & $-0.01748(6)$ & $-0.01551(4)$ & $-0.01235(3)$ \\
6 & $\;\;\>0.00756(6)$ & $\;\;\>0.01191(8)$ & $\;\;\>0.01257(6)$ & $\;\;\>0.01067(6)$ & $\;\;\>0.00816(4)$ \\
7 & $-0.00607(8)$ & $-0.01003(7)$ & $-0.0106(1)$ & $-0.00936(9)$ & $-0.00737(7)$ \\
8 & $\;\;\>0.0051(1)$ & $\;\;\>0.0088(2)$ & $\;\;\>\underline{0.0100(1)}$ 
& $\;\;\>\underline{0.00902(8)}$ & $\;\;\>\underline{0.00722(7)}$ \\
9 & $-0.00427(8)$ & $-0.0083(1)$ & $-0.0100(1)$ & $-0.0095(2)$ & $-0.00762(8)$ \\
10 & $\;\;\>0.00381(8)$ & $\;\;\>0.0082(2)$ & $\;\;\>0.0106(2)$ & $\;\;\>0.0106(2)$ & \\
11 & $-0.00309(3)$ & $-0.0075(1)$ & $-0.0109(2)$ & $-0.0116(2)$ & \\
12 & $\;\;\>0.00259(3)$ & $\;\;\>0.00728(7)$ & $\;\;\>0.0115(2)$ & $\;\;\>0.0136(3)$ & \\
13 & $-0.00220(3)$ & $-0.0071(3)$ & $-0.0125(3)$ & $-0.0156(3)$ & \\
14 & $\;\;\>0.00188(4)$ & $\;\;\>0.0068(4)$ & $\;\;\>0.0140(5)$ & $\;\;\>0.0172(4)$ & \\
15 & $-0.00151(5)$ & $-0.0060(5)$ & $-0.0145(7)$ & $-0.0227(7)$ & \\
16 & $\;\;\>0.00131(6)$ & & & & \\
17 & $-0.00096(8)$ & & & & \\
\hline
\end{tabular}
\end{center}
\normalsize
\end{table}
 
\begin{table}[!htbp]
\caption[]{${\mathsf R \left(\insiva\right)}/B_2^{k-1}={\mathsf
R}_{4,1}/B_2^{k-1}.$ The underline marks the approximate location of
the minimum value.}
\label{ivatab}
\vspace{0.2cm}
\footnotesize
\begin{center}
\begin{tabular}{|l|l|l|l|l|}
\hline
$k$ & \multicolumn{1}{c|}{$D = 2$} & \multicolumn{1}{c|}{$D = 3$} & \multicolumn{1}{c|}{$D = 4$} & \multicolumn{1}{c|}{$D = 5$} \\
\hline
5 & & $-0.000242(2)$ & $-0.000550(2)$ & $-0.000765(4)$ \\
6 & & $\;\;\>0.000435(3)$ & $\;\;\>0.000984(7)$ & $\;\;\>0.00137(1)$ \\
7 & $-1.03(5)\times10^{-8}$ & $-0.000342(5)$ & $-0.00078(1)$ & $-0.00112(1)$ \\
8 & $\;\;\>5.0(2)\times10^{-7}$ & $\;\;\>0.000294(6)$ 
& $\;\;\>0.00076(1)$ & $\;\;\>\underline{0.00107(1)}$ \\
9 & $-3.5(2)\times10^{-6}$ & $-0.000290(6)$ &
$\underline{-0.00073(1)}$ 
& $-0.00108(2)$ \\
10 & $\;\;\>8.6(4)\times10^{-6}$ & $\;\;\>\underline{0.000278(6)}$ 
& $\;\;\>0.00074(1)$ & $\;\;\>0.00121(2)$ \\
11 & $-1.53(8)\times10^{-5}$ & $-0.000281(6)$ & $-0.00082(2)$ & $-0.00135(3)$ \\
12 & $\;\;\>1.83(9)\times10^{-5}$ & $\;\;\>0.000273(8)$ & $\;\;\>0.00085(2)$ & $\;\;\>0.00154(3)$ \\
13 & $-1.90(9)\times10^{-5}$ & $-0.00029(1)$ & $-0.00090(3)$ & $-0.00178(4)$ \\
14 & $\;\;\>2.1(1)\times10^{-5}$ & $\;\;\>0.00028(2)$ & $\;\;\>0.00104(4)$ & $\;\;\>0.00205(5)$ \\
15 & $-1.79(9)\times10^{-5}$ & $-0.00026(2)$ & $-0.00102(6)$ & $-0.00244(9)$ \\
16 & $\;\;\>1.6(1)\times10^{-5}$ & & & \\
17 & $-1.3(1)\times10^{-5}$ & & & \\
18 & $\;\;\>1.3(2)\times10^{-5}$ & & & \\
\hline
\end{tabular}
\end{center}
\normalsize
\end{table}
 
\begin{table}[!htbp]
\caption[]{${\mathsf R \left(\insve\right)}/B_2^{k-1}={\mathsf R}_{5,5}/B_2^{k-1}$}
\label{vdtab}
\vspace{0.2cm}
\footnotesize
\begin{center}
\begin{tabular}{|l|l|l|l|}
\hline
$k$ & \multicolumn{1}{c|}{$D = 2$} & \multicolumn{1}{c|}{$D = 3$} & \multicolumn{1}{c|}{$D = 4$} \\
\hline
6 & $\;\;\>0.00848(7)$ & $\;\;\>0.01328(9)$ & $\;\;\>0.01286(5)$ \\
7 & $-0.0063(1)$ & $-0.0094(1)$ & $-0.00947(9)$ \\
8 & $\;\;\>0.00469(9)$ & $\;\;\>0.0078(1)$ & $\;\;\>0.0081(1)$ \\
9 & $-0.00363(7)$ & $-0.0065(1)$ & $-0.00748(9)$ \\
10 & $\;\;\>0.00290(6)$ & $\;\;\>0.0061(1)$ & $\;\;\>0.0076(1)$ \\
11 & $-0.00230(8)$ & $-0.0056(1)$ & $-0.0072(1)$ \\
12 & $\;\;\>0.0020(1)$ & $\;\;\>0.0051(1)$ & $\;\;\>0.0075(2)$ \\
13 & $-0.0017(2)$ & $-0.0045(2)$ & $-0.0080(2)$ \\
14 & $\;\;\>0.0013(2)$ & $\;\;\>0.0048(3)$ & $\;\;\>0.0077(3)$ \\
15 & $-0.00097(24)$ & $-0.0043(4)$ & $-0.0087(5)$ \\
\hline
\end{tabular}
\end{center}
\normalsize
\end{table}
 
\begin{table}[!htbp]
\caption[]{$\mathrm{Pinwheel}/B_2^{k-1}$}
\label{pinwheeltab}
\vspace{0.2cm}
\footnotesize
\begin{center}
\begin{tabular}{|l|l|l|l|}
\hline
$k$ & \multicolumn{1}{c|}{$D = 2$} & \multicolumn{1}{c|}{$D = 3$} & \multicolumn{1}{c|}{$D=4$} \\
\hline
5 & $-0.0266(1)$ & $-0.0315(1)$ & $-0.02647(6)$ \\
6 & $\;\;\>0.00906(9)$ & $\;\;\>0.01000(8)$ & $\;\;\>0.00754(6)$ \\
7 & $-0.00194(4)$ & $-0.00230(3)$ & $-0.00178(3)$ \\
8 & $\;\;\>0.000187(4)$ & $\;\;\>0.000302(6)$ & $\;\;\>0.000261(5)$ \\
9 & $-3.7(1)\times 10^{-6}$ & $-0.0000157(5)$ & $-0.000022(1)$ \\
10 & $\;\;\>2(1)\times 10^{-8}$ & $\;\;\>5(1)\times 10^{-7}$ & $\;\;\>8(4)\times 10^{-7}$ \\
\hline
\end{tabular}
\end{center}
\normalsize
\end{table}

\clearpage

\section{Graphs of the ratio of selected diagrams to high order.}
\label{highordgraphs}

In this appendix we graph the ratios $B_k[4,1]/B_k[0,1]$,
$B_k[5,1]/B_k[0,1]$, ${\mathsf{R}}_{4,2}/{\mathsf{R}}$,
${\mathsf{R}}_{4,1}/{\mathsf{R}}$, and ${\mathsf{R}}_{5,5}/{\mathsf{R}}$ to
orders up to $k=17$.

\begin{figure*}[!htb]
\begin{picture}(100,65)(0,0)%
\put(0,40){${B_k[4,1]}/{B_k[0,1]}$}
\centerline{\includegraphics[scale=0.35,origin=c,angle=-90]{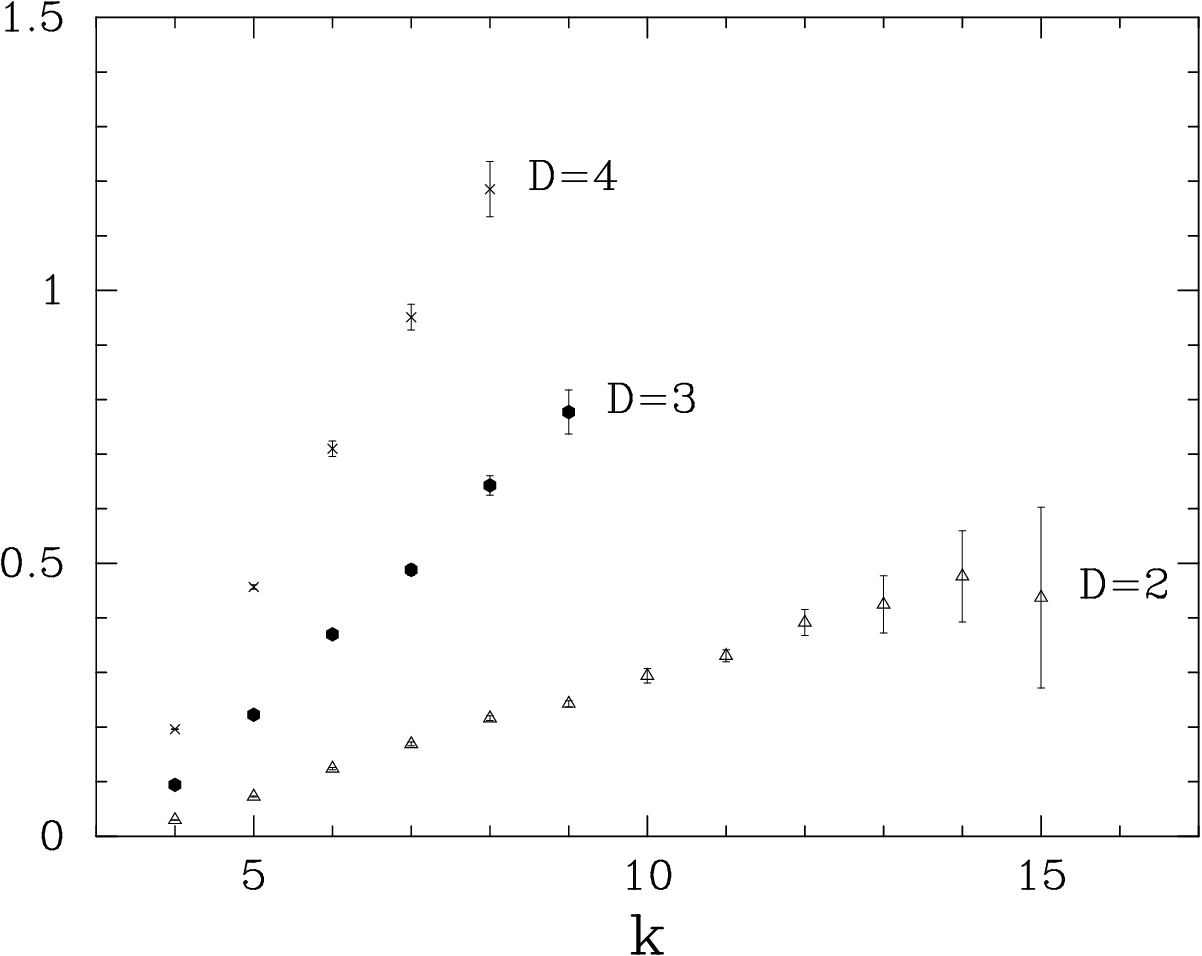}}
\end{picture}%
\caption[]{Absolute value of $\wigglyb/\wigglya=B_k[4,1]/B_k[0,1]$ in dimensions 
2 (triangles), 3 (filled circles),
and 4 (crosses).}
\label{rh2graph}
\end{figure*}

\begin{figure*}[!htb]
\begin{picture}(100,65)(0,0)%
\put(0,37){${B_k[5,1]}/{B_k[0,1]}$}
\centerline{\includegraphics[scale=0.35,origin=c,angle=-90]{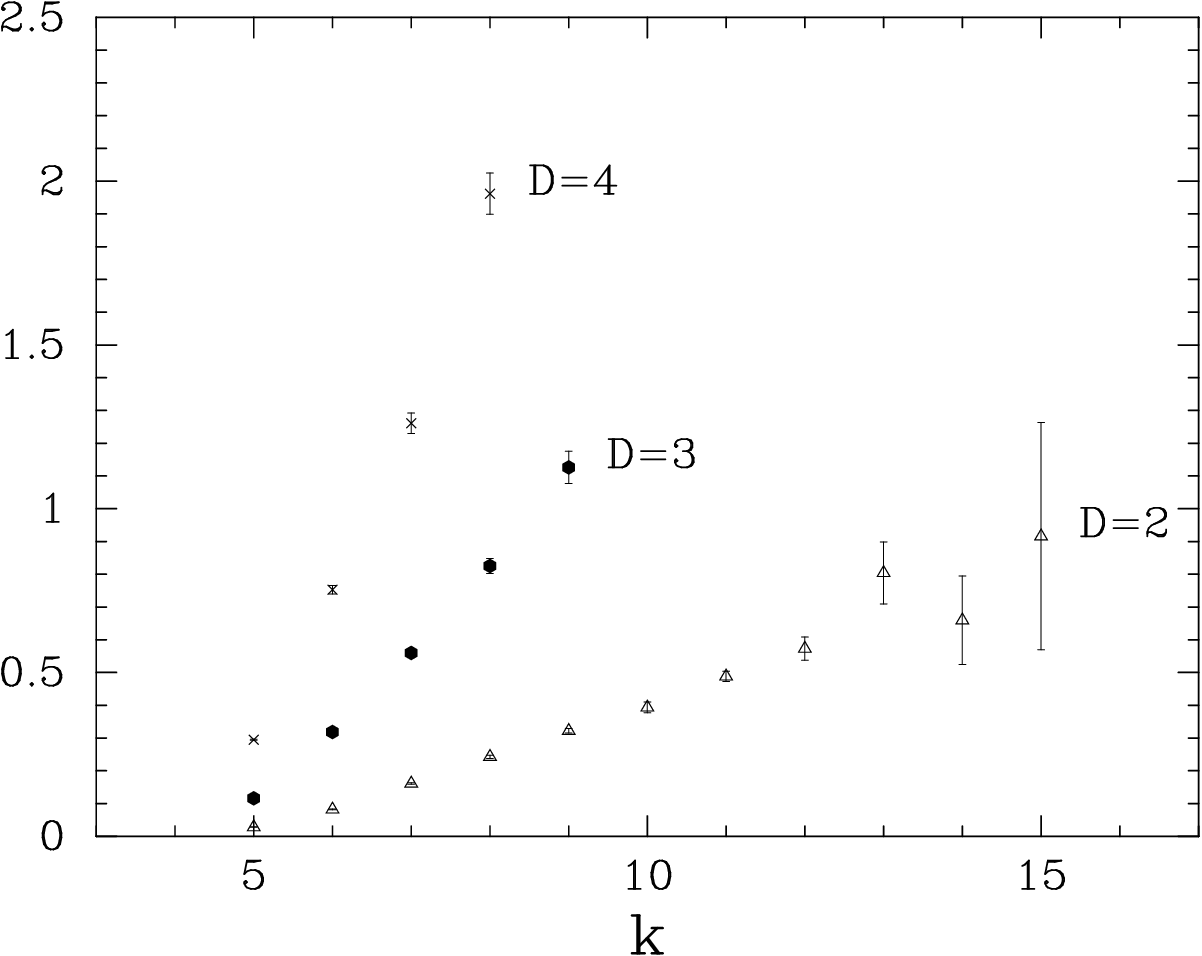}}
\end{picture}%
\caption[]{Absolute value of ${\wigglyc}/\wigglya=B_k[5,1]/B_k[0,1]$ in dimensions 2
(triangles), 
3 (filled circles),
and 4 (crosses).}
\label{rh3graph}
\end{figure*}

\begin{figure*}[!htb]
\begin{picture}(100,65)(0,0)%
\put(10,40){${{\mathsf R}_{4,2}}/{{\mathsf R}}$}
\centerline{\includegraphics[scale=0.35,origin=c,angle=-90]{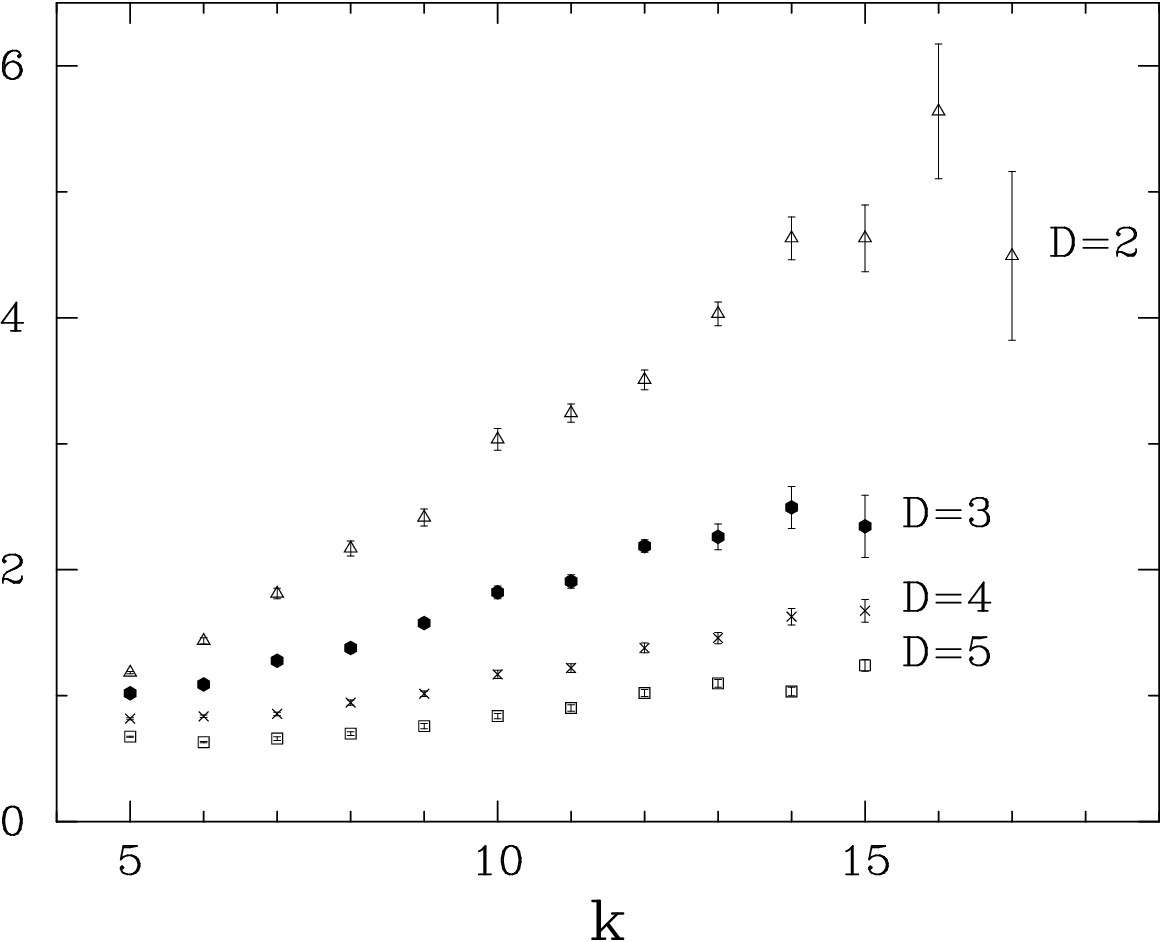}}
\end{picture}%
\caption[]{Absolute value of ${\mathsf R}(\insivb)/R={\mathsf R}_{4,2}/{\mathsf R}$ in dimensions 2 (triangles), 3 (filled circles), 4 (crosses), and 5 (squares).}
\label{loose4bgraph}
\end{figure*}

\begin{figure*}[!htb]
\begin{picture}(100,65)(0,0)%
\put(10,40){${{\mathsf R}_{4,1}}/{{\mathsf R}}$}
\centerline{\includegraphics[scale=0.35,origin=c,angle=-90]{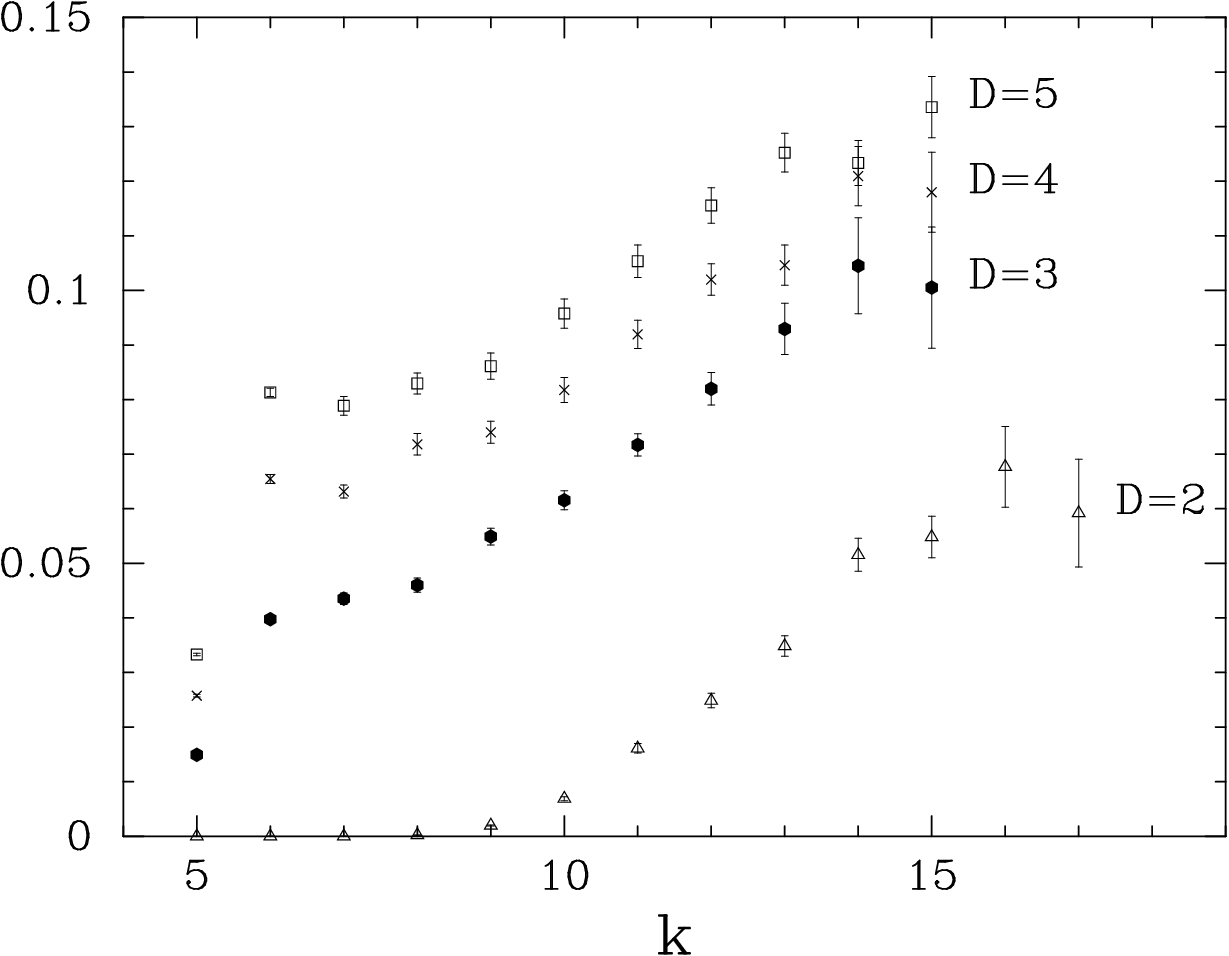}}
\end{picture}%
\caption[]{Absolute value of ${\mathsf R}(\insiva)/R={\mathsf R}_{4,1}/{\mathsf R}$ in dimensions 2 (triangles), 3 (filled circles), 4 (crosses), and 5 (squares).}
\label{loose4agraph}
\end{figure*}

\begin{figure*}[!htb]
\begin{picture}(100,65)(0,0)%
\put(10,40){${{\mathsf R}_{5,5}}/{{\mathsf R}}$}
\centerline{\includegraphics[scale=0.35,origin=c,angle=-90]{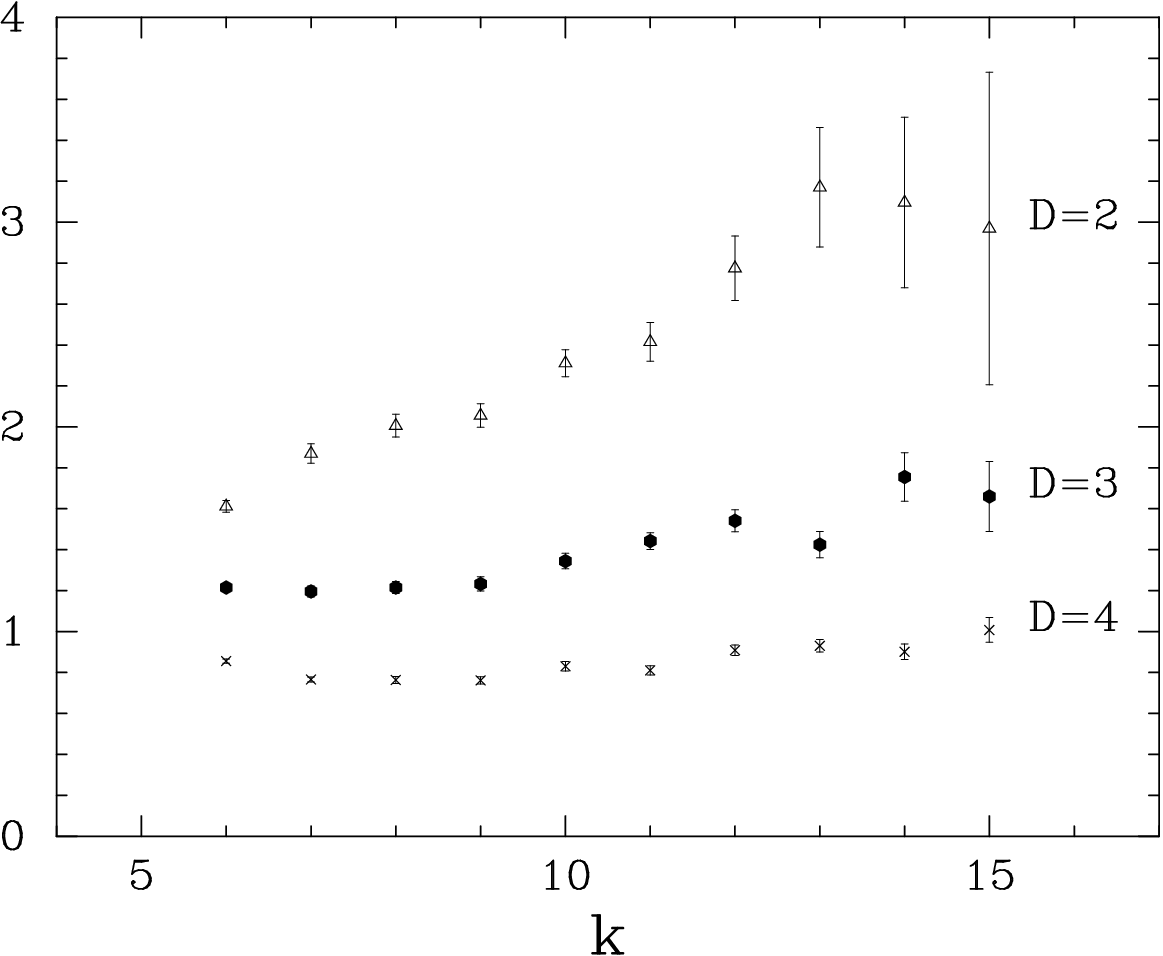}}
\end{picture}%
\caption[]{Absolute value of ${\mathsf R}(\insve)/{\mathsf R}={\mathsf R}_{5,5}/{\mathsf R}$ 
in dimensions 2 (triangles), 3 (filled circles),
and 4 (crosses).}
\label{loose5egraph}
\end{figure*}

\clearpage


\noindent {\bf Acknowledgments:} {This work was supported in part by 
the National Science 
Foundation under DMR-0073058. We thank Prof.~R.~J.~Baxter and
Prof.~G.~Stell for useful discussions.}


\end{document}